\documentclass[a4paper, USenglish, cleveref, autoref, thm-restate]{lipics-v2021}
\usepackage[utf8]{inputenc}
\hideLIPIcs
\nolinenumbers

\usepackage{graphicx}
\usepackage[disable]{todonotes}
\usepackage{amssymb}
\usepackage{amsmath}
\usepackage{amssymb}
\usepackage{amsfonts}
\usepackage{amstext}
\usepackage{amsthm}
\usepackage{mathtools}
\usepackage{xspace}
\usepackage{xparse}

\definecolor{codegreen}{rgb}{0,0.6,0}
\definecolor{codegray}{rgb}{0.5,0.5,0.5}
\definecolor{codepurple}{rgb}{0.58,0,0.82}
\definecolor{backcolour}{rgb}{0.95,0.95,0.92}

\Crefname{figure}{Fig.}{Figs.}

\newcommand{\MS}{{\texttt{MS-algorithm}}}
\newcommand{\NEW}{{\texttt{NEW-algorithm}}} 

\newcommand{\NCE}{E_{nc}}

\NewDocumentCommand{\cvector}{m o}{%
	\IfNoValueTF{#2}
	{\ensuremath{\mathbf{#1}}\xspace}
	{\ensuremath{\mathbf{#1#2}}\xspace}
}

\NewDocumentCommand{\nvector}{m o}{%
	\IfNoValueTF{#2}
	{\ensuremath{\widehat{\mathbf{#1}}}\xspace}
	{\ensuremath{\widehat{\mathbf{#1#2}}}\xspace}
}

\title{An algorithm for accurate and simple-looking metaphorical maps}

\author{Eleni Katsanou}{National Technical University of  Athens,  Greece} {ekatsanou@mail.ntua.gr}{https://orcid.org/0000-0002-1001-1411}{}
\author{Tamara Mchedlidze}{Utrecht University, Utrecht, Netherlands}{t.mtsentlintze.uu.nl}{https://orcid.org/0000-0001-6249-3419}{}
\author{Antonios Symvonis}{National Technical University of  Athens,  Greece}{symvonis@math.ntua.gr}{https://orcid.org/0000-0002-0280-741X}{}
\author{Thanos Tolias}{National Technical University of  Athens,  Greece}{thanostolias@mail.ntua.gr}{https://orcid.org/0009-0003-8354-8855}{}

\authorrunning{E. Katsanou, T. Mchedlidze, A. Symvonis, T. Tolias}

\Copyright{Eleni Katsanou, Tamara Mchedlidze, Antonios Symvonis, Thanos Tolias} 

\relatedversion{This is the extended version of E. Katsanou, T. Mchedlidze, A. Symvonis, T. Tolias, "An algorithm for accurate and simple-looking metaphorical maps'', to appear in the Proc. of the 33rd International Symposium on Graph Drawing and Network Visualization, GD 2025, LIPIcs, Volume 357, 2025.}



\begin{document}

\maketitle

\begin{abstract}
\emph{Metaphorical maps} or \emph{contact representations} are visual representations of vertex-weighted graphs that rely on the geographic map metaphor. The vertices are represented by countries, the weights by the areas of the countries, and the edges by contacts/boundaries among them. The \emph{accuracy} with which the weights are mapped to areas and the \emph{simplicity} of the polygons representing the countries are the two classical optimization goals for metaphorical maps.
    Mchedlidze \& Schnorr~\cite{MchedSchnorr22} presented a force-based algorithm that creates metaphorical maps that balance between these two optimization goals. Their maps look visually simple, but the accuracy of the maps is far from optimal -- the countries' areas can vary up to 30\% compared to required.
    In this paper, we provide a multi-fold extension of  the algorithm in~\cite{MchedSchnorr22}. More specifically:
    \begin{enumerate}
        \item Towards improving accuracy: We introduce the notion of region stiffness and suggest a technique for varying the stiffness based on the current pressure of map regions.
        \item Towards maintaining simplicity: We introduce a weight coefficient to the pressure force exerted on each polygon point based on whether the corresponding point appears along a narrow passage.
        \item Towards generality: We cover, in contrast to ~\cite{MchedSchnorr22}, non-triangulated graphs. This is done by either generating points where more than three regions meet or by introducing holes in the metaphorical map.  
    \end{enumerate}
    We perform an extended experimental evaluation that, among other results, reveals that our algorithm is able to construct metaphorical maps with nearly perfect area accuracy with a little sacrifice in their simplicity.  
\end{abstract}

\section{Introduction}

\begin{figure}[t]
	\centering
    \begin{minipage}[t]{0.48\textwidth}
    \centering
    \includegraphics[width=0.8\linewidth]{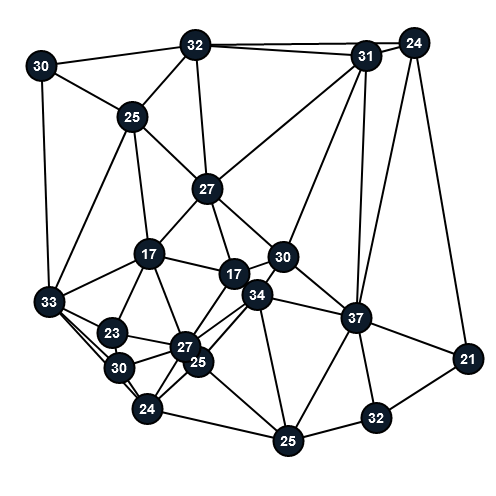}
    \end{minipage}
    \begin{minipage}[t]{0.48\textwidth}
    \centering
    \includegraphics[width=0.8\linewidth]{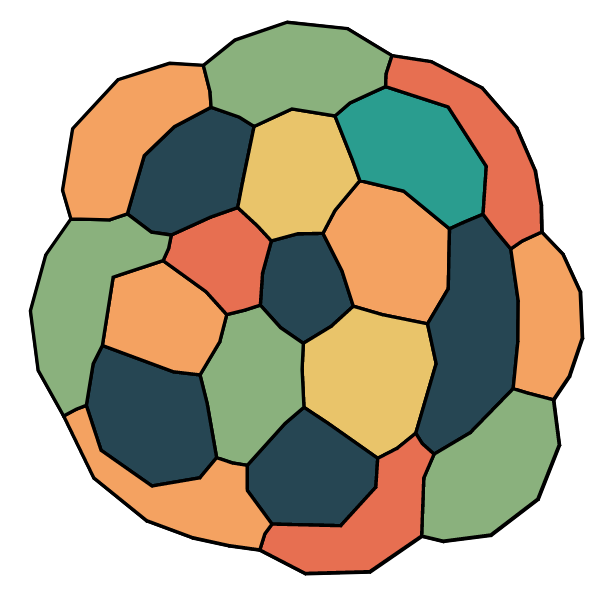}
    \end{minipage}
    \caption{An internally triangulated plane graph with vertex weights and its corresponding metaphorical map.}
    \label{fig:example_metaphoricMap}
\end{figure}

Metaphorical maps or map-like graph visualization is an alternative way to the popular
node-link diagrams~\cite{2013gd} and matrix representations~\cite{WuTC2008} for representing graphs.
In these visualizations vertices are represented by polygonal regions and edges by contacts among them; see~\cref{fig:example_metaphoricMap}. Metaphorical maps have their individual advantage: they rely on the human familiarity with geographic maps and therefore instantly spark curiosity in the viewers by looking fairly familiar. It has been experimentally confirmed that map-like visualizations of graphs are more enjoyable than node-link diagrams~\cite{SaketSK16}.  
They also outperform treemaps in the tasks that require recognition of hierarchy~\cite{Biuk-AghaiPP17}. 
Another important advantage of these visualizations is that they can naturally display the weights associated with the graph's vertices, by the mean of resizing the map's regions. Such visualizations are also known as \emph{area-proportional contact representations}~\cite{thesisAlam15} and are closely related to cartograms~\cite{tobler2004thirty,nusrat2016state}, visualizations created by deforming geographic maps with a goal to represent values of a variable (e.g. population, number of votes) by the sizes of the deformed countries.  Hogr{\"a}fer, Heitzler, and Schulz ~\cite{HograferHS20} presented a unifying framework, showing that map-like representations of graphs and cartograms lie on two opposite sides of a single spectrum of map-like visual representations.

Map-like visual representations have been studied for more than 50 years~\cite{tobler2004thirty} and a few excellent surveys have been devoted to this topic~\cite{HograferHS20,nusrat2016state,tobler2004thirty}. Generally, the construction of map-like representations is guided by the following criteria.  \emph{Statistical accuracy} refers to how well the modified areas represent the corresponding statistic shown; with the \emph{cartographic error} measuring by how much the actual area of a region is away from the desired. \emph{Geographical accuracy} refers to how much the modified shapes and locations of the regions resemble those in the original map.  Preserving geographical accuracy is a goal in many algorithms generating cartograms. However, for metaphorical maps, the geographical accuracy is not relevant as there is no given geography to preserve.
Finally, \emph{topological accuracy} refers to how well the topology of the cartogram matches the topology of the original map. %
In case of map-like representations of planar graphs topological accuracy must be fully preserved -- all edges of the given planar graph must be represented as regions' contacts and each region contact must correspond to an edge.

A criterion that is less explicitly mentioned in the map-like representation literature is the complexity of the outlines of map's regions. In general, keeping the region outlines simple is desired, since one of the tasks the users of these visualizations face, is the comparison of the region areas. The user performance in this task differs even when circles are compared to rectangles~\cite{David96:seeing}. Motivated by this, a few works concentrated on generating map-like representations with simple regions.  For instance, the Dorling cartogams~\cite{Dorling} use circles as regions, and Demers cartograms~\cite{BortinsDC} use square regions. Also, mosaic cartograms \cite{CanoBCPSS15} tend to have simpler looking outlines of regions, since they are constituted by segments having a limited amount of slopes. Contrary to that, the approaches where no care is taken of the complexity of the regions, tend to produce regions with very complex outlines, refer to~\cite{nusrat2016state}. 
 
 In theory research, the complexity of regions was formalized as the maximum number of sides per region. A series of works was devoted to reducing this metric for rectilinear area-proportional contact representation from the initial 40~\cite{BergMS09} to the final 8~\cite{alam2013linear}.
 Other relevant series of work on \emph{area-universal planar drawings}, starts with a planar drawing and an assignment of weights to its faces. The requirement is to modify the drawing, by preserving its planar embedding and realize the desired face areas. 
Thomassen~\cite{thomassen1992plane} showed that every plane cubic graph is \emph{area-universal} (perfect cartographic accuracy is achievable) and additionally in the resulting map the faces are bounded by triangles. Kleist~\cite{kleist2018drawing} showed that 1-subdivision of any plane graph is area-universal. However, all the theoretical approaches that guarantee a few sides per region and perfect cartographic accuracy~\cite{alam2013linear,kleist2018drawing,thomassen1992plane} are not guaranteed from having very narrow pointed areas in the regions and therefore the diagrams do not really look simple; refer for instance to \cref{fig:example_Alam_20}.(b,f)
.  

Motivated by this challenge, Mchedlidze and Schnorr~\cite{MchedSchnorr22} used a more sophisticated complexity measure introduced in~\cite{BrinkhoffKSB95} to evaluate the quality of the polygons in their metaphorical maps, that is able to capture the intuitive perception of a polygon's complexity. They presented a force-based algorithm that creates metaphorical maps that balance between the two optimization goals: cartographic error and polygon complexity. Their maps look visually simple, but the cartographic error can be up to 30\%.

\subparagraph{Our 
contribution.}
The known techniques  either build  complex metaphorical maps with perfect cartographic accuracy~\cite{alam2013linear,kleist2018drawing,thomassen1992plane}  or simple-looking metaphorical maps with cartographic error up to 30\%~\cite{MchedSchnorr22}. In this work, we address this gap and aim towards producing simple metaphorical  maps with near zero cartographical error. 
\begin{itemize}
\item We present a multi-fold extension of  the algorithm in~\cite{MchedSchnorr22}. More specifically:
    \begin{itemize}
        \item Towards improving accuracy: We introduce the notion of region stiffness and suggest a technique for varying the stiffness based on the current pressure of map regions.
        \item Towards maintaining simplicity: We introduce a weight coefficient to the pressure force exerted on each polygon point based on whether the corresponding point appears along a narrow passage.
        \item Towards generality: We cover, in contrast to ~\cite{MchedSchnorr22}, non-triangulated graphs. This is done by either generating points where more than three regions meet or by introducing holes in the metaphorical map.
   \end{itemize}
\item We perform an extended experimental evaluation which aims  to:
\begin{itemize}
\item Compare our algorithm to the algorithm of Mchedlidze and Schnorr~\cite{MchedSchnorr22} with respect to cartographic accuracy and complexity.
\item %
Examine the cartographic accuracy vs the regions' complexity trade-off.
\item Evaluate the extension to non-triangulated graphs.
\item Evaluate the influence of the initial layouts on the quality of the maps, by comparing the initial layout as suggested in~\cite{MchedSchnorr22} to the cartographically accurate initial layout given by the theoretical approach in~\cite{AlamBFKKT2013}. 
\end{itemize}
\item Our experiments among other facts indicate that the suggested algorithm achieves almost perfect cartographic accuracy with only a small increase in polygon complexity compared to~\cite{MchedSchnorr22}.
\end{itemize}

\textbf{Paper organization.} In \cref{sec:Preliminaries}, we present our notation, the quality metrics employed in evaluating cartographic maps as well as the basic ingredients of the algorithm of  Mchedlidze and Schnorr~\cite{MchedSchnorr22}. In \cref{sec:our_algorithm}, we present our algorithm which   (a) introduces the notion of region stiffness, (b) applies corrective weight coefficients on pressure forces, and (c)    handles non-triangulated plane graphs. Our experimental evaluation is presented in \cref{sec:ExpermentalEvaluation}. Finally, we conclude in~\cref{sec:future_work} with directions for future work.%

A demo application for generating metaphorical maps has been implemented in JavaScript using yFiles \cite{yFiles2025} and is available as a web application at \url{http://aarg.math.ntua.gr/demos/metaphoric_maps/}. All experiments were executed using a Java implementation. To allow replicability, we make the implementation and the whole graph benchmark publicly available at \url{https://github.com/ekatsanou/metaphorical-maps}.

\section{Preliminaries}
\label{sec:Preliminaries}

In this section, we introduce the notation used throughout the paper and we present the quality metrics of cartographic accuracy and polygon complexity that are employed in our experimental evaluation. Given that the algorithm presented in this paper is an extension of the algorithm of Mchedlidze and Schnorr~\cite{MchedSchnorr22} (shortly \MS), we also present its brief description. 

Let $G=(V, E, w)$ be a vertex-weighted graph where $w:V \rightarrow \mathbb{R^+}$ is its weight function. In a \emph{metaphorical map} $M=\mu(G)$ of $G$,  vertex $v\in V$ is depicted as polygonal \emph{region} (\emph{country}) $\mu(v)$ so that adjacent vertices share a non-trivial contact (\emph{boundary}). Our optimization goal is to build metaphorical maps such that the area of the region $\mu(v)$, denoted by $A(\mu(v))$, is roughly equal to $w(v)$. 

In a metaphorical map $M(G)$, we refer to the points that define its polygonal regions also as \emph{vertices}. Given two vertices $u,~v$ of the metaphorical map, we denote by $\cvector{u}[v]$ the vector from $u$ to $v$.

\subsection{Quality measures for Metaphorical Maps}
As mentioned in the introduction, we measure the quality of a metaphorical map by its cartographic accuracy -- for this, similarly to the previous work, e.g.~\cite{alam2013linear,Alam2015,MchedSchnorr22}, we employ the metric \emph{normalized cartographic error}. The simplicity of the map is measured by the metric \emph{polygon complexity}, introduced in~\cite{BrinkhoffKSB95} and applied for the evaluation of metaphorical maps in~\cite{MchedSchnorr22}.

\subparagraph{Normalized cartographic error}
Consider a vertex $v\in V$ of a vertex-weighted graph $G=(V,E,w)$ and its corresponding region $\mu(v)$ in its metaphorical map $\mu(G)$. The \emph{normalized area of region $\mu(v)$}, denoted by $A^\prime(\mu(v))$, is defined as:
\begin{equation*}
A^\prime(\mu(v))  
 \coloneqq A(\mu(v)) \cdot \frac{\sum\limits_{u \in V}{w(u)}}{\sum\limits_{u \in V} {A(\mu(u))}}
\end{equation*}

\noindent Then, the \emph{normalized cartographic error $\NCE$ of region $\mu(v)$} is defined as:

\begin{equation}
\NCE(\mu(v))  \coloneqq \frac{ \mid A^\prime(\mu(v)) - w(v) \mid }{\max\{A^\prime(\mu(v)),w(v)\}}
\label{eq:normalized_charto_error}
\end{equation}

\subparagraph{Polygon Complexity}
Brinkhoff, Kriegel, Schneider, and Braun~\cite{BrinkhoffKSB95}  defined the complexity of a polygon $P$ as a function of three quantities (see also Appendix~\ref{app:complexity}), that can be intuitively understood based on an example of a star-shape:  (a) the \emph{frequency of $P$'s vibration}, denoted by $\text{freq}(P)$,  which specifies how many tips a star-shape has,  (b) the \emph{amplitude} of $P$'s vibration, denoted by $\text{ampl}(P)$, which specifies how long the tips of a star-shape are and (c) $P$'s \emph{convexity}, denoted by $\text{conv}(P)$, the fraction of the area that is not covered within the smallest enclosing circle. The \emph{polygon complexity} is then defined as 
$\text{compl}(P) =
0.8 \cdot \text{ampl}(P) \cdot \text{freq}(P) + 0.2 \cdot \text{conv}(P)$, which maps $P$ to values in $\lbrack0,1\rbrack$, with low values meaning low complexity.      

In our experiments, we quantify the quality of a map by both the average and the maximum value of normalized cartographic error and polygon complexity over the regions of the map.

\subsection{The \MS}
\label{subsec:MSalgo}
The \MS~\cite{MchedSchnorr22} is a typical force-directed algorithm that employs several antagonistic forces applied to the vertices of the metaphorical map, that, hopefully, at  equilibrium produce a layout with good characteristics, i.e.,  low %
normalized cartographic error and low %
polygon complexity. The algorithm employs three forces (vertex-vertex repulsion, vertex-edge repulsion, and angular-resolution) targeted towards producing metaphorical maps of low polygon complexity and one force (air-pressure) which works towards reducing the cartographic error of the metaphorical map.
Here, we only describe the air-pressure force, since our proposed algorithm modifies how this force is applied. The exact definition of the other forces can be found in  Appendix~\ref{app:forces} and in~\cite{MchedSchnorr22}. 

The \emph{normalized air pressure} in region $g$, $P(g)$, exerts a force on each bounding edge $e$ based on the pressure's magnitude and the edge's length $\ell(e)$ in relation to the length of the entire polygonal region boundary $\text{circ}(g)$; see \cref{app:forces} for a formal definition of $P(g)$. 
Thus,  the \emph{air-pressure force} on edge $e$ of region $g$ was defined as  $\cvector{F}(e,g) =3  P(g)\frac{\ell(e)}{\text{circ}(g)}\nvector{r}$,  
where $\nvector{r}$ is a unit vector perpendicular to $e$ directed towards outside of $g$. Force $\cvector{F}(e,g)$ was applied to both endpoints of edge $e$.

The force-directed algorithm is run for $iter$ number of steps, the value of this parameter is determined experimentally. 
Notice that the resultant of the forces applied to a vertex of the map, may force it to cross over an edge, and thus, the displacement of each vertex must be limited to prevent this from happening. To achieve that, the \MS~ adopted the rules of ImPrEd~\cite{SimonettoAAB2011impred} that ensure the preservation of the planar embedding of the map.  It should be also noted that the constant factors employed in each of the described forces have been determined experimentally.  

A final note regarding the number of vertices defining each polygonal region.
This number fluctuates during the execution of the algorithm in order to allow each region to obtain a more elaborate (or simpler) shape. Let $\bar{\ell}$ denote the average edge length over all edges in the metaphorical map. Provided that no edge crossings are introduced, the \MS~ eliminates vertices of degree 2 that become  closer than $\frac{1}{10}\bar{\ell}$ to their neighbor and it splits in half the edges that get longer than  $2\bar{\ell}$.

\todo[inline]{****REVIEWER***: Is it better to use the term \emph{point} and \emph{segment} when we refer to polygons and keep the term vertices and edges for graphs only?
Eleni: Polygon points is not very well defined. How would we describe a point that lies inside the polygon, or a point that lies on an edge of the polygon?\\
subdivision points???}

Each force-directed algorithm requires for its execution an initial layout. Given a plane vertex-weighted graph $G$, the \MS~  constructs an initial metaphorical map by considering the dual of a planar drawing of $G$. The algorithm is designed for internally triangulated graphs. The dual vertices of the inner faces are placed in the barycenters of the triangles representing those faces. The dual edges are drawn as polylines consisting of two segments meeting at a bend-point, which lies in the middle of the corresponding primal edge. Details can be found in~\cite{MchedSchnorr22}, refer also to~\cref{fig:transformation} in the Appendix.

\section{Our Metaphorical Map Generation Algorithm}
\label{sec:our_algorithm}
In this section, we describe our extension of the \MS. More specifically, we describe how to revise the air-pressure force utilized in the \MS~  by (a) incorporating a  \emph{stiffness coefficient}  
and (b) by applying a corrective \emph{weight coefficient} that aims to eliminate narrow passages in the metaphorical map.  The constants involved in the calculations of our algorithm are determined experimentally. Finally, we show how the algorithm is adapted in order to accommodate non-triangulated graphs.  

\subparagraph{The Stiffness of each Map Region}
After examining the output of the \MS, we observed that some regions had noticeably higher cartographic error than others, leading to the conclusion that some attributes of a region made it less responding to air pressure and, thus, resulted in high cartographic errors. 
Therefore, for every region we introduce a variable, which we refer to it as \emph{stiffness coefficient}, that accounts for its stiffness. Let $g$ be an internal region and let $P_i(g)$ be its pressure at the $i$-th iteration of the force-directed algorithm. If at iteration $i$ region $g$ is over-pressured (i.e., $P_i(g)>1$) we increase the region's stiffness coefficient by a small amount $step$; if it is under-pressured (i.e., $P_i(g)<1$) we decrease it by $step$, otherwise, it remains unchanged. In addition, we restrict the stiffness coefficient in the range 
$\lbrack s_{\textit{low}} , s_{\textit{high}}\rbrack$ 
where $s_{\textit{high}} \geq 1$ and 
$s_{\textit{low}}= \frac{1}{s_{\textit{high}}}$. 
An appropriate value for $s_{\textit{high}}$ 
is determined experimentally. 

Given a region $g$, we define the \emph{stiffness coefficient of $g$ at the $i$-th iteration of the algorithm}, denoted by $s_i(g)$, as:

\begin{equation*}
\begin{cases} 
s_{0}(g) = 1 \\
s_i(g) = \text{min}(s_{\textit{high}}, \text{max}(s_{\textit{low}} , s_{i-1}(g) + \alpha \cdot \textit{step})), ~~~~~~i>0
\end{cases}
\end{equation*}

where  \(\alpha =  \begin{cases} 
      -1, & P_{i-1}(g) < 1 \\
      0, & P_{i-1}(g) = 1 \\
      1, & P_{i-1}(g) > 1 
   \end{cases}
\)

Then, the \emph{revised air-pressure force} of  region $g$ on its boundary edge $e$ during the $i$-th iteration of the algorithm, $i\geq 1$, denoted by $\cvector{F^\prime_i}(e,g)$ is defined as 
\begin{equation}
\cvector{F^\prime_i}(e,g) =  s_i(g) \cdot \cvector{F_i}(e,g) 
\end{equation}
where  $\cvector{F_i}(e,g)$ is the air-pressure force computed at the $i$-th iteration of  the \MS. Note that the ``stiffness'' attribute is different for each region and demonstrates an adaptive behaviour over time. Employing such adaptive coefficients in force/energy-directed drawing algorithms appears to be uncommon, mainly due to performance issues. However, we note that a similar scheme has been employed in the work of Hu~\cite{Hu05} (referred to as \emph{adaptive cooling scheme}).   

We performed a few experiments (refer to Appendix~\ref{app:parameters}) with 20-node graphs in order to determine the parameters of the algorithm, that resulted in the following values: $\textit{step}=0.02$, $\textit{iter}=1.000$ and $s_\textit{high}=8$. To also account for the size of the input graph, assuming that larger graphs take a longer time to converge to a good metaphorical map, we decided to set the number of iterations to $\textit{iter}= 800 + 10n$, the value consistent with $\textit{iter}=1.000$ for $n=20$. 

\subparagraph{Improving the Visual Complexity}
When we applied the stiffness coefficients to the regions, we observed that they tended to create long, narrow passages, thereby reducing the visual complexity of the metaphorical map.  To mitigate this effect, we introduce a new corrective weight coefficient for the pressure forces exerted on each regionÔÇÖs vertices, while keeping the total pressure of the region constant.  These weight coefficients redistribute pressure so that vertices in narrow passages receive a larger share of the force, whereas vertices in wider parts of the region receive less.
Before defining the weight coefficient, we introduce some auxiliary notation:
\begin{itemize}
	\item Let $M$ be a metaphorical map, and let
	$ \rho(M) = \sqrt{\left(\sum\limits_{g \in M} {A(g)} \right) /\pi} $
	denote the radius of a circle whose area equals the total area of $M$.  We refer to $\rho(M)$ as the \emph{ideal radius} of $M$.
	
	\item Let $g$ be a region of $M$,  let $u$ be one of its vertices and let $e$ be an edge of $g$ that is not adjacent to $u$.  Further, let $x$ be the point of $e$ that is closest to $u$. The Euclidean distance of $u$ to $e$, denoted by  $d_E(u, e)$, is simply  the Euclidean distance from $u$ to point $x$.  
    We also define the \emph{polygonal distance} of vertex $u$ to edge $e$, denoted by $d_P(u,e)$, as the length of the shortest path when moving from $u$ to $x$ on the boundary of $g$. 	
\end{itemize}

\noindent
Intuitively, an edge $e$ is sufficiently ``opposite'' to a vertex $u$ if it is close to it and, at the same time, it  belongs to the opposite side of a narrow passage. For each vertex $u$ of region $g$, we want to identify the edge $e$ of $g$ that is the closest out of those opposite to it.  In order to  avoid selecting edges that are collinear with $u$, we  consider only edges for which it holds that $d_E(u,e) < 0.9\,d_P(u,e)$ and, out of those, we select the edge $e$ of smallest Euclidean distance to $u$. We refer to this edge $e$ as the \emph{pairing edge} of $u$ and we denote it by $pair(u)$. Further, we denote the Euclidean distance from $u$ to $pair(u)$   by $d(u,g)$.

Consider the metaphorical map $M$. We introduce a scale factor based on the ideal radius $\rho(M)$.  We regard $0.05\,\rho(M)$ as the minimum ``acceptable'' passage width and,   hence,  we define the quantity $\delta(u, g) = \frac{0.05\,\rho(M)}{d(u, g)}$ which assumes values greater than 1 when $u$ is very close to its pairing edge $pair(u)$.

We now define  the corrective coefficient $\beta(u,g)$ that will be applied to the air-pressure forces at each vertex $u$ of region $g$ of the metaphorical map as:  
\[
\beta(u, g)=1 + \operatorname{sign}\bigl(\delta(u,g) - 1\bigr)\ln\!\Bigl(1 + \bigl|\delta(u,g) - 1\bigr|\Bigr).
\]

\noindent
Hence:
\begin{itemize}
	\item If $\delta(u,g) = 1$, i.e., $u$ marginally does not participate in a narrow passage, then $\beta(u,g) = 1$.
	\item If $\delta(u,g) > 1$, then $\beta(u,g) = 1 + \ln\bigl(\delta(u,g)\bigr),$ which grows logarithmically once $\delta(u,g)$ exceeds 1.
	\item If $0 < \delta(u,g) < 1$, then $\beta(u,g) = 1 - \ln\bigl(2 - \delta(u,g)\bigr),$ which remains below 1 but varies smoothly as $\delta(u,g)$ decreases.
\end{itemize}

Let $u_i$ and $u_{i+1}$ be the endpoints of edge $e_i$ of region $g$ and let $\nvector{r}$ be a unit vector perpendicular to $e_i$ directed towards the exterior of $g$.
Recall that, the air-pressure force on $e_i$ is defined as  $\cvector{F}(e_i,g) =3  P(g) s(g) \frac{\ell(e_i)}{\text{circ}(g)}\nvector{r} $, where $s(g)$ is the stiffness coefficient in the current iteration.
Since the air-pressure force is  exerted on both endpoints of $e_i$, the total air-pressure force on $e_i$ is $2\times3P(g)s(g) \frac{\ell(e_i)}{\text{circ}(g)}\nvector{r} $ and the total air-pressure force of region $g$ is $2\times3P(g)s(g)$. 

The new air-pressure force, employing the above introduced corrective coefficient $\beta( \cdot,g)$,  on edge $e_i=(u_i,u_{i+1})$ is now defined as
$\cvector{F}(e_i, g) = 
$ \[ \underbrace{ 3P(g)s(g)\frac{2\beta(u_i,g)\ell(e_i)}{\sum_j{\ell(e_j)(\beta(u_j,g)+ \beta(u_{j+1},g))}}\nvector{r}}_{\text{Force applied on $u_i$}}
+ \underbrace{ 3P(g)s(g)\frac{2\beta(u_{i+1},g)\ell(e_i)}{\sum_j{\ell(e_j)(\beta(u_j,g)+ \beta(u_{j+1},g))}}\nvector{r}}_{\text{Force applied exerted on $u_{i+1}$}}.\]

\noindent
Therefore, the total air-pressure force applied to region $g$ equals to  
$\sum_i \cvector{F}(e_i, g)$. So, we have  that: 

\[\sum_i \cvector{F}(e_i, g) = \frac{6P(g)s(g)}{\sum_j{\ell(e_j)(\beta(u_j,g)+ \beta(u_{j+1},g))}} \sum_i{(\beta(u_i,g)\ell(e_i)+\beta(u_{i+1},g)\ell(e_i))}\] $= 2 \times 3P(g)s(g)$,
and thus,  the total air-pressure force on region $g$ remains unchanged.

\subsection{Dealing with non-Triangulated Graphs}
In this section, we discuss how to create %
metaphorical maps for biconnected, non-triangulated plane graphs. Mchedlidze and Shnorr~\cite{MchedSchnorr22} focused on the case of internally triangulated graphs. However, their algorithm was, in principle, able to also deal with non-triangulated graphs provided that, in the initial layout of the given graph, the \emph{barycenter visibility property} is satisfied, that is, the barycenter of each non-triangulated face is located within that face and all vertices of the face are \emph{visible} from the barycenter, that is, the open segment that connects the barycenter and any vertex of the face does not cross the boundary of the face.

\begin{figure}[tb]
    \centering
    \begin{minipage}[t]{.24\textwidth}
    \centering
    \includegraphics[width=1\linewidth]{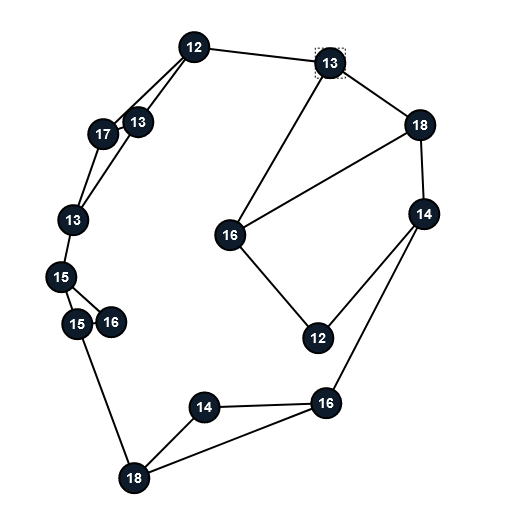}
    \subcaption{}
    \end{minipage}
    \begin{minipage}[t]{.24\textwidth}
    \centering
    \includegraphics[width=1\linewidth]{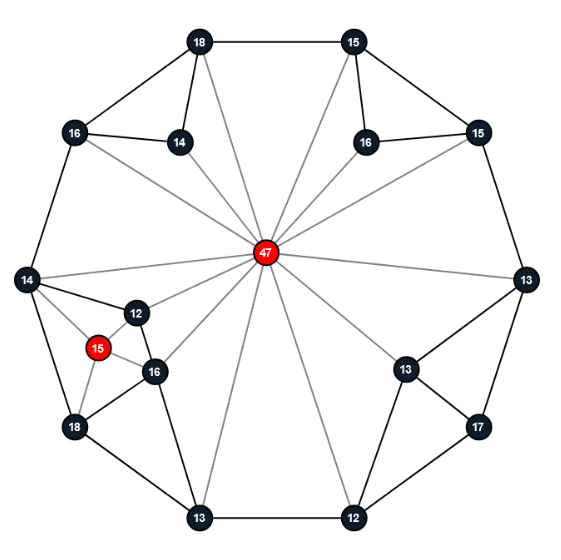}
    \subcaption{\nolinenumbers}
    \end{minipage}
    \begin{minipage}[t]{.24\textwidth}
    \centering
    \includegraphics[width=1\linewidth]{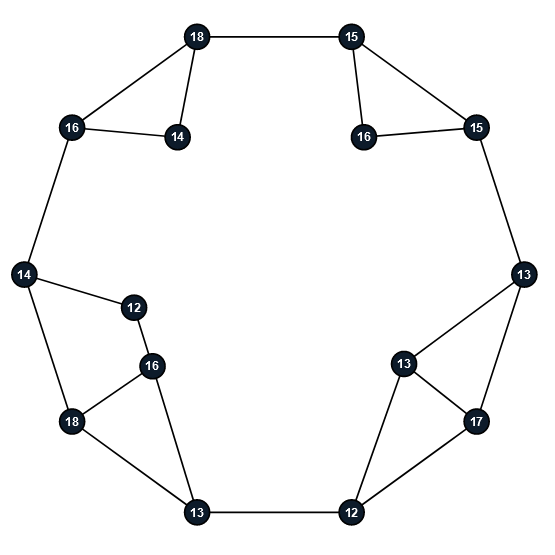}
    \subcaption{\nolinenumbers}
    \end{minipage}
    \newline
    \begin{minipage}[t]{.24\textwidth}
    \centering
    \includegraphics[width=0.9\linewidth]{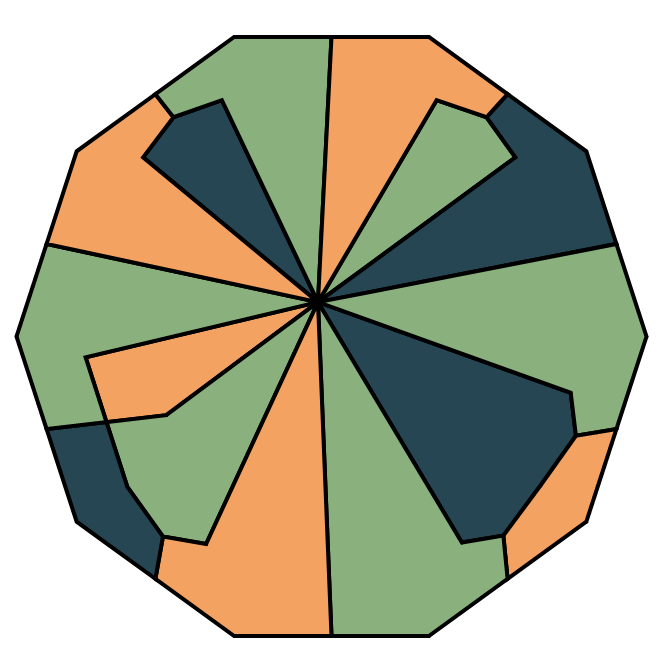}
    \subcaption{}
    \end{minipage}
    \begin{minipage}[t]{.24\textwidth}
    \centering
    \includegraphics[width=0.9\linewidth]{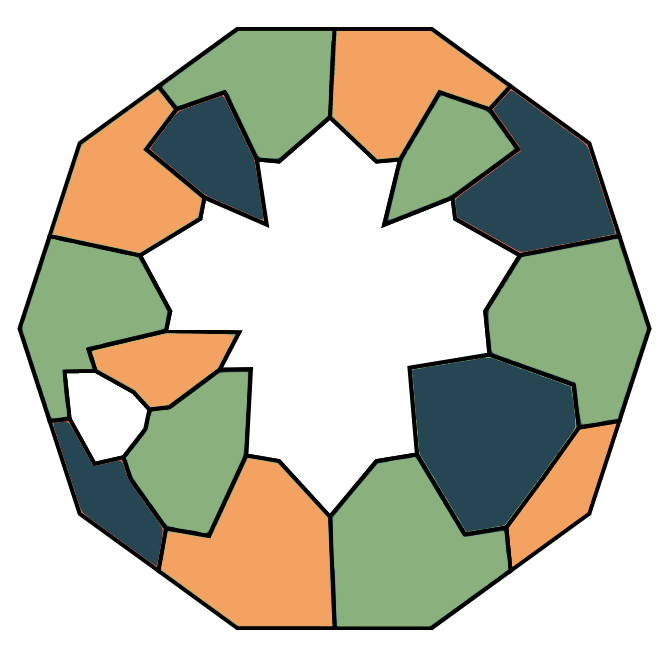}
    \subcaption{\nolinenumbers}
    \end{minipage}
    \begin{minipage}[t]{.24\textwidth}
    \centering
    \includegraphics[width=0.9\linewidth]{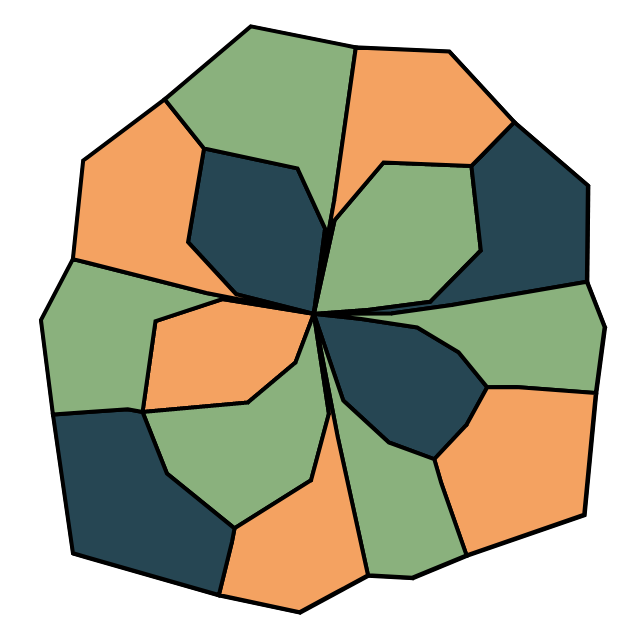}
    \subcaption{\nolinenumbers}
    \end{minipage}
    \begin{minipage}[t]{.24\textwidth}
    \centering
    \includegraphics[width=0.9\linewidth]{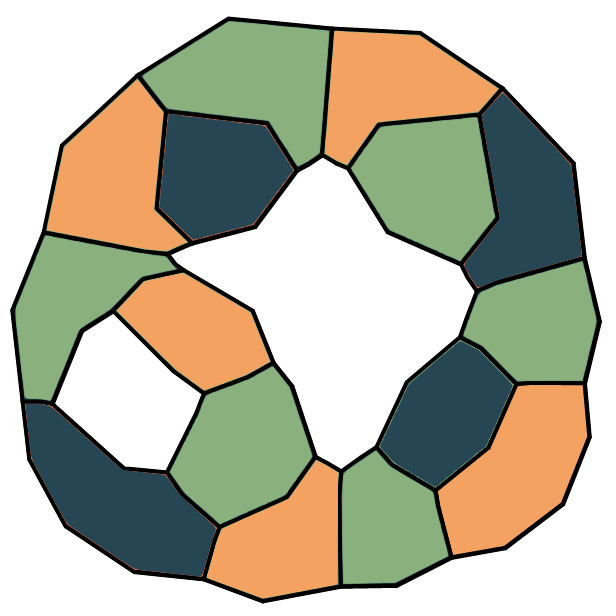}
    \subcaption{\nolinenumbers}
    \end{minipage}
    \caption{Construction of the initial layout for non-triangulated graphs. (a) A graph $G$ with a large non-triangulated face. (b) The auxiliary graph $G'$ -- red vertices are added, drawn with the Tutte's algorithm. (c) The auxiliary vertices are removed. (d) Transformation of the drawing in~\cref{fig:non_triangulated_trans}.(c) based on the graph's dual (as employed in the \MS). (e) Transformation of the drawing in~\cref{fig:non_triangulated_trans}.(c) based on the graph's dual  and by treating the auxiliary vertices as holes. (f-g) Forces are applied to ~\cref{fig:non_triangulated_trans}.(d) and ~\cref{fig:non_triangulated_trans}.(e), respectively; holes are weighted according to the proposed function.\label{fig:non_triangulated_trans}}
\end{figure}

We proceed to establish the barycenter visibility property as follows. %
We internally triangulate the graph by creating a new \emph {auxiliary} vertex for each non-triangulated inner face and connecting it to every vertex of the face. 
The use of auxiliary vertices for mesh generation, often called Steiner points, is a common technique to transform general grids into high-quality triangulations (see \cite{BernE1995, Owen1998ASO}).
Let $G'= (V \cup V_{aux}, E \cup E_{aux})$  be the resulting internally triangulated graph where $V_{aux}$ and $E_{aux}$ are the sets of auxiliary vertices and the extra edges used in the triangulation, respectively.  
We apply on $G'$  Tutte's barycentric embedding algorithm~\cite{Tutte1963} that fixes the outerface on a circle and places every inner vertex on the barycenter of its neighbors, resulting in a planar straight line layout  $\Gamma_{bar}(G')$; refer to \cref{fig:non_triangulated_trans}.(a-b)
We finish by removing the auxiliary vertices and their adjacent edges from $\Gamma_{bar}(G')$; refer to \cref{fig:non_triangulated_trans}.(c). It is trivial to see that the derived layout satisfies the barycenter visibility property. Indeed, the auxiliary vertex of each non-triangulated face $f$ is placed at the barycenter of its neighbors, that is,  the vertices of $f$. Therefore, in the constructed planar drawing, face $f$ has its barycenter lying inside it. Moreover, the fact that the auxiliary edges inside $f$ do not cross the boundary of $f$, ensures the visibility between the barycenter of $f$ and all of  $f$'s vertices. 

After ensuring that the barycenter of every inner face lies within its boundary, we are in a position to obtain an initial metaphorical map by using the transformation deployed in the \MS; refer to Section~\ref{subsec:MSalgo} and \cref{fig:non_triangulated_trans}.(d). Note that this transformation leads to points in the metaphorical maps where more than three regions meet. Recall that in our definition of the metaphorical map, only the non-degenerate contacts represent adjacencies, therefore the maps constructed this way still represent exactly the adjacencies present in the given graph.

\subparagraph{Metaphorical maps with holes.}

\begin{figure}[t]
\centering  
\begin{minipage}[t]{.48\textwidth}
    \centering
    \includegraphics[width=0.6\textwidth]{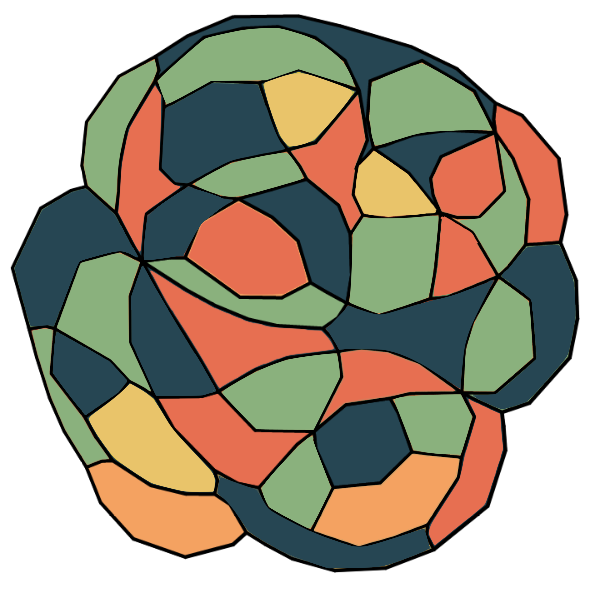}
    \subcaption{Map with high degree contact points.}
\end{minipage}
\begin{minipage}[t]{.48\textwidth}
    \centering
     \includegraphics[width=0.6\textwidth]{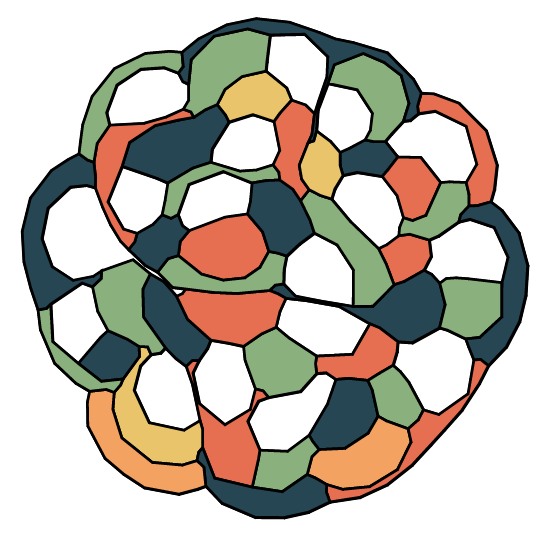}
    \subcaption{\nolinenumbers Map with holes.}
\end{minipage}
\caption{Two maps which correspond to the same 40-vertex graph. Non-triangulated regions are represented in (a) with high degree contact points and in (b) with holes.}
\label{fig:no_holes_problem}
\end{figure}

By visually inspecting the maps created with the above method, we noticed that the large faces (in the initial graph) which result in multiple point contacts (in the map) appear to be quite cluttered; refer to \cref{fig:non_triangulated_trans}.(d) and \cref{fig:no_holes_problem}.(a). Therefore, we  propose an alternative method that treats large faces as holes; Refer to~\cref{fig:non_triangulated_trans}.(e). Practically, obtaining such maps is very simple -- we can just consider the graph $G'$ obtained after the introduction of the auxiliary vertices, transform it to a map, as suggested in the \MS~, and treat regions corresponding to the auxiliary vertices as holes. When applying forces to the map vertices, we can treat holes in exactly the same way we treat other regions. 
However, since the holes do not have a target weight, we have the flexibility to assign those weights with the goal to improve the  appearance of the overall map. Our intuition is that the more regions are adjacent to a hole, and the heavier those regions are, the bigger the hole needs to be. 
After some experimentation, we found that the following  function for assigning weights to  holes produces reasonably good results, however,  with further testing better functions may arise:
\begin{equation}
\label{eq:holeweight}
   \textit{holeWeight}(h_i) = \frac{1}{4 \text{deg}(h_i)} \left[ \sum\limits_{(v_j, h_i) \in E(G)} ^{j}\sqrt{w(v_j)}\right]^2
\end{equation}
Note that, we do not prioritize minimizing the normalized cartographic error within holes, as they do not represent regions. However, it is important that holes remain visually simple to preserve the clarity of adjacencies. To allow adjacent regions the flexibility to expand naturally, we do not apply stiffness coefficients to holes. Nonetheless, we do apply the corrective weight coefficients of the pressure forces to maintain overall layout stability.

\section{Experimental Evaluation}
\label{sec:ExpermentalEvaluation}
In our experimental evaluation we aim to answer the following research questions:
\begin{enumerate}
    \item What is the performance of our algorithm compared to the performance of the \MS~ based on the quality metrics of normalized cartographic error and polygon complexity? Does the difference in performance depend on the input characteristics such as number of vertices, nesting ratio or weight ratio (refer to Section~\ref{sec:testBed} for the definitions)? 
    \item Is there a trade-off between cartographic error and polygon complexity? How do we control it?
    \item Does  the initial metaphorical map  have an effect on the quality of the final one?
    \item Is the handling of non-triangulated biconnected graphs satisfactory? 
    \item Is the proposed algorithm practical? Does its running time scale well with the size of the input graph?
\end{enumerate} 
The analysis of question 5 
can be found in Appendix~\ref{app:evaluationplus}. 

\subparagraph{The Graph Test Data}
\label{sec:testBed}
For the generation of test data, we followed the approach of \cite{MchedSchnorr22} (described in detail in  Algorithm 5.1 of \cite{Schnorr2020Thesis} with a uniform distribution).
The authors of~\cite{MchedSchnorr22} generate planar straight-line drawings based on the Delaunay triangulation of a random point set. A fraction $\textit{nest} \in [0, 1]$ of the points, called \emph{nesting ratio},  is placed within the triangles of the initial Delaunay triangulation. It was experimentally verified in~\cite{MchedSchnorr22} that the nesting ratio has an effect on the algorithm's performance, therefore, we also included it in our analysis.
 For every vertex,  we generate a weight using a uniform distribution with different \emph{weight ratios} of maximum to minimum weight. 
The above described procedure is augmented to generate non-triangulated plane graphs as follows.  Following the generation of an internally triangulated plane graph, we remove a number of randomly selected internal edges in order to create non-triangulated plane graphs. We denote the ratio of removed to initial internal edges by $rem$. Note that an edge is only removed if the graph remains biconnected. 

\subparagraph{Comparison to the \MS}

\begin{figure}[tb]
	\begin{minipage}[b]{.9\linewidth}
		\centering 
		\includegraphics[page=1]{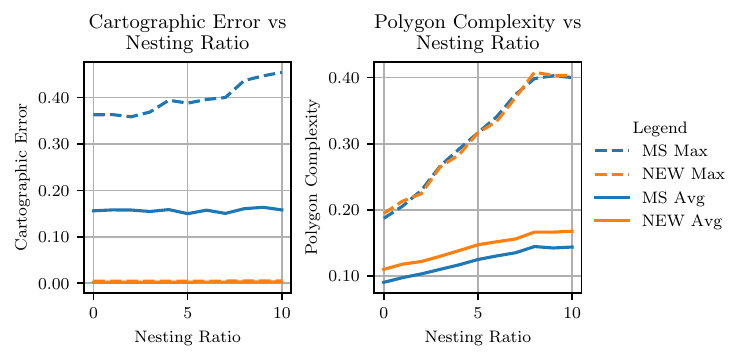} 
	\end{minipage}
	\caption{\NEW~vs  \MS~ for different values of nesting ratio.}
    \label{fig:nesting_ratio}
\end{figure}

\begin{figure}[tb]
	\begin{minipage}[b]{.9\linewidth}
		\centering 
		\includegraphics[page=1]{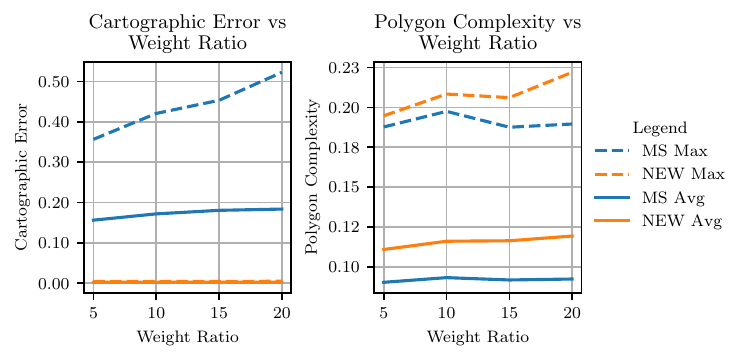} 
	\end{minipage}
	\caption{\NEW~vs  \MS~ for different values of weight ratio.}
    \label{fig:weight_ratio}
\end{figure}

\begin{figure}[tb]
	\begin{minipage}[b]{.9\linewidth}
		\centering 
		\includegraphics[page=1]{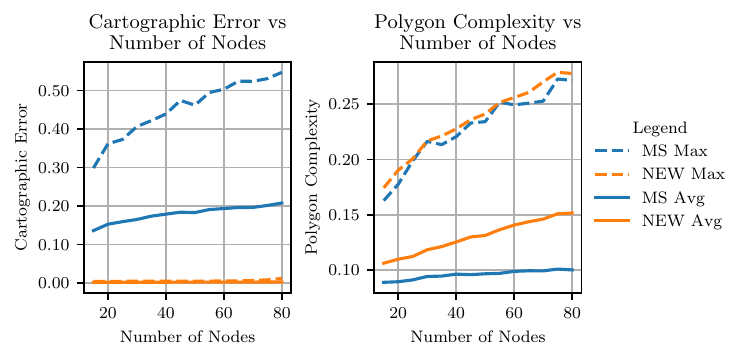} 
	\end{minipage}
	\caption{\NEW~vs  \MS~ for different number of nodes.}
	\label{fig:number_of_nodes}
\end{figure}

For the comparison of the performance of our algorithm (\NEW) against \MS, we examined how three parametersÔÇöthe nesting ratio $nest$ (see \cref{fig:nesting_ratio}), the weight ratio $w$ (see \cref{fig:weight_ratio}), and the total number of nodes $n$ (see \cref{fig:number_of_nodes})--influence both algorithms. To this end, we conducted three experiments:

\begin{enumerate}
	\item \textit{Varying nesting ratio.} For each value of $nest \in \{0, 0.1, 0.2, \dots, 1.0\}$, we generated 50 graphs with $n = 20$ nodes and weight ratio $w = 5$.
	
	\item \textit{Varying weight ratio.} For each $w \in \{5, 10, 15, 20\}$, we generated 50 graphs with $n = 20$ nodes and nesting ratio $nest = 0$.
	
	\item \textit{Varying number of nodes.} For each $n \in \{15, 20, 25, \dots, 80\}$, we generated 50 graphs with nesting ratio $nest = 0$ and weight ratio $w = 5$ and we let the algorithms run for $800 + 10n$ iterations.
\end{enumerate}

\begin{figure}[htb]
\centering  
\begin{minipage}[t]{.24\textwidth}
    \centering
    \includegraphics[width=0.88\linewidth]{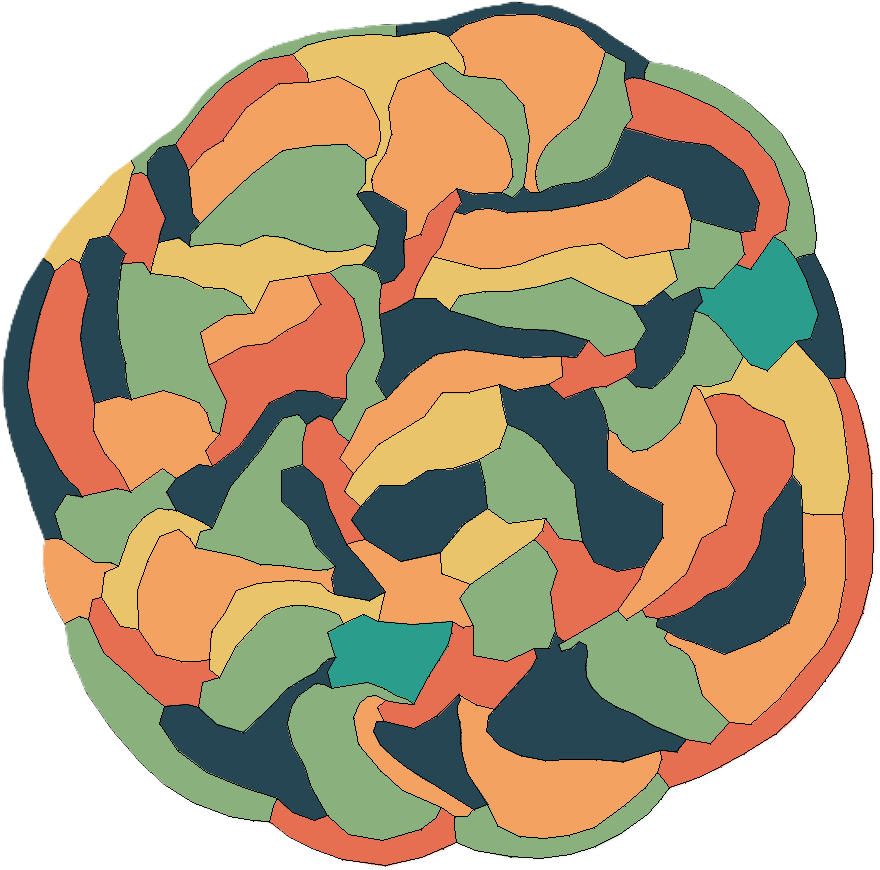}
    \subcaption{\NEW}
\end{minipage}
\begin{minipage}[t]{.24\textwidth}
    \centering
    \includegraphics[width=0.88\linewidth]{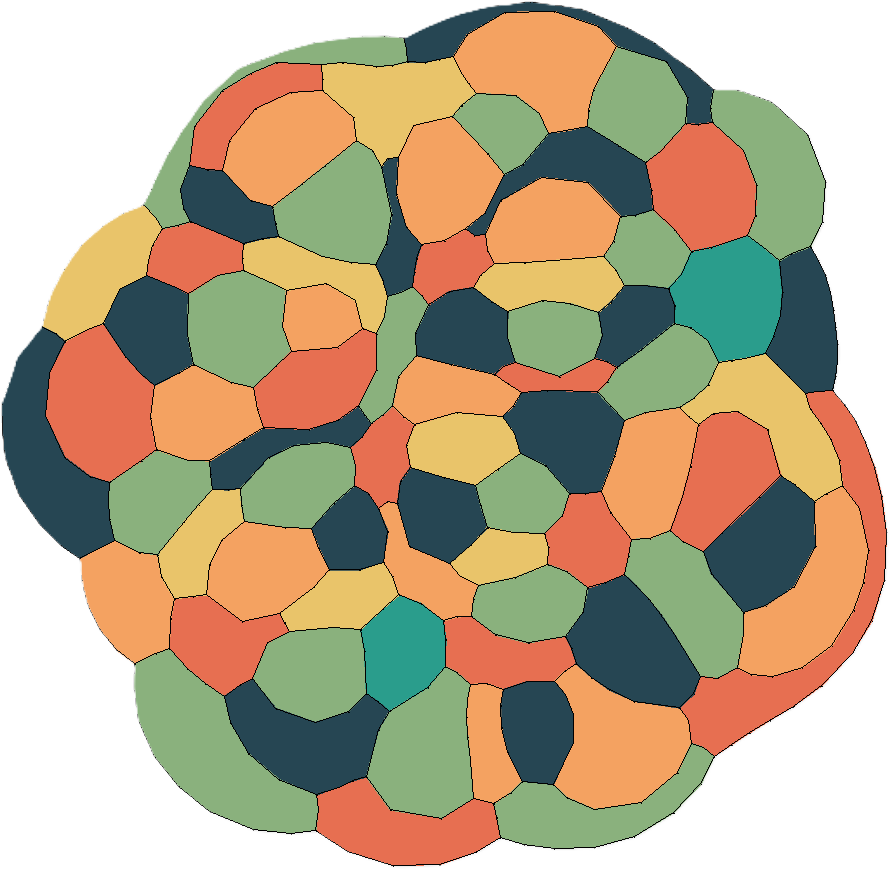}
    \subcaption{\nolinenumbers \MS}
\end{minipage}
\begin{minipage}[t]{.24\textwidth}
    \centering
    \includegraphics[width=0.88\linewidth]{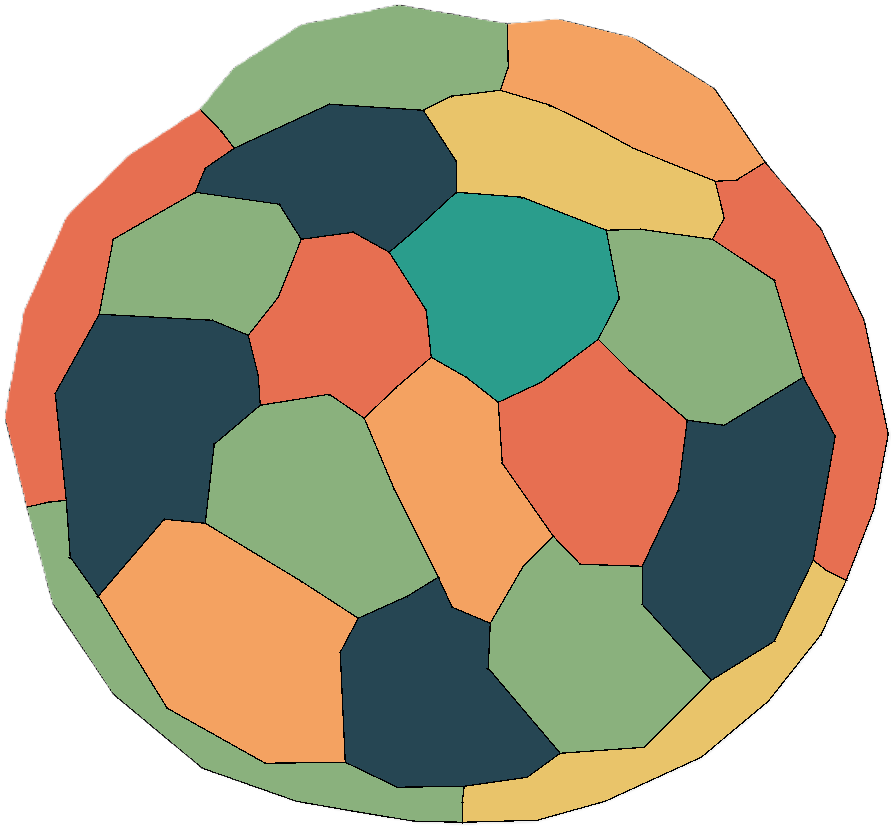}
    \subcaption{\nolinenumbers \NEW}
\end{minipage}
\begin{minipage}[t]{.24\textwidth}
    \centering
    \includegraphics[width=0.88\linewidth]{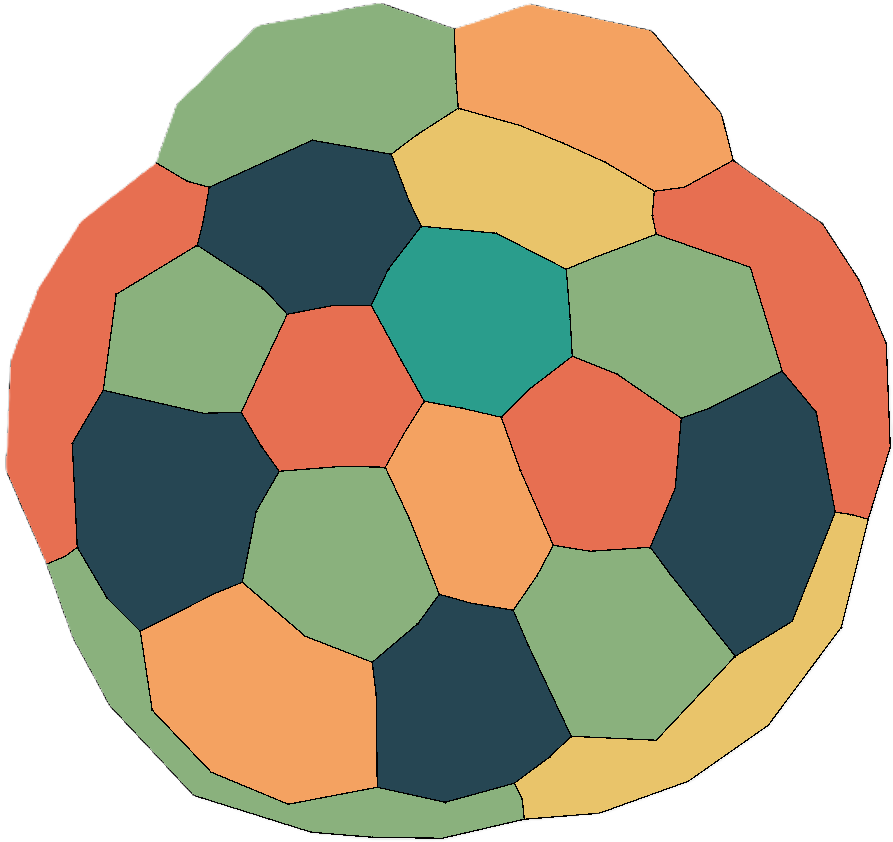}
    \subcaption{\nolinenumbers \MS}
\end{minipage}
\caption{(a-b) The maps that correspond to the largest increase in average polygon complexity. The corresponding graph has 75 vertices. (c-d) Maps for a random graph. }
\label{fig:mapexamplesWorse}
\end{figure}

The results of every comparison follow a similar trend. Our algorithm achieves average cartographic error close to zero at the cost of a  slightly higher polygon complexity. In all of our experiments (1450 in total), the recorded average normalized cartographic error (for each metaphorical map created) was in the range $\lbrack 0.12 \%, 0.44 \%\rbrack$. In comparison, the same metric %
for \MS~ is in the range $\lbrack 8.8 \% , 24.7 \%\rbrack$. The price we paid was a small increase in the average polygon complexity. The highest observed increase in average polygon complexity over all the maps was $6.9 \%$ and was observed for $n=75$, where the maximum value of average polygon complexity was $19.6 \%$, well below the $40\%$ value where polygons are considered to be complex~\cite{BrinkhoffKSB95}. The maps in our experiment that correspond to this maximum value of increase in the average polygon complexity are shown in~\cref{fig:mapexamplesWorse}.(a-b). We observe that indeed our map's regions are longer and, sometimes, narrower. However, in the case of 
smaller (or ``simpler'') 
graphs, this increase in complexity is hardly noticeable, see e.g.~\cref{fig:mapexamplesWorse}.(c-d). 
Regarding the properties of the graphs, all three parameters (nesting ratio, weight ratio, and the number of vertices) affect the polygon complexity of our produced maps.

\subparagraph{Cartographic error and polygon complexity trade-off}
\label{sec:carError_PolCompl_tradeoff}

\begin{figure}[htb]
\centering  
\begin{minipage}[t]{0.95\linewidth}
    \centering
    \includegraphics[page=1]{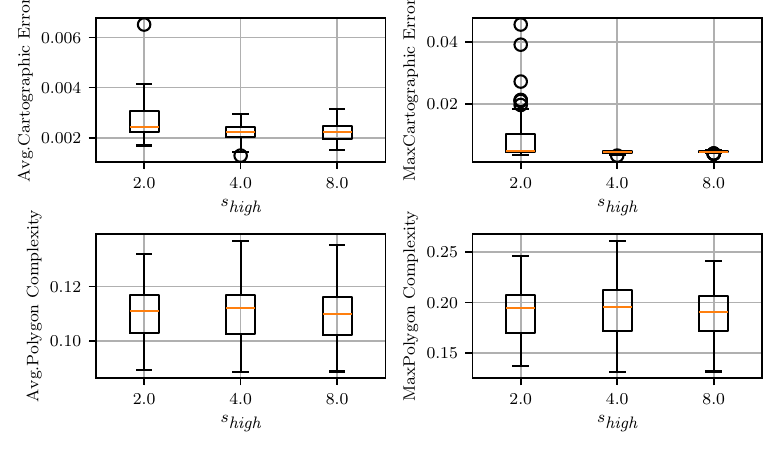}
\end{minipage}
\caption{Cartographic Error and Polygon Complexity for different values of maximum stiffness, $n = 20$, $w = 5$, $\textit{nest} = 0$.}
\label{fig:maxStiffness}
\end{figure}

The cost we pay for achieving average cartographic error close to zero is a small increase in the average polygon complexity. Given that our algorithm is identical to the \MS~ for stiffness
$s_{\textit{high}}=1 $ (without the application of  the corrective weight coefficients on the pressure forces), we evaluated our algorithm for several values of $s_{\textit{high}} \in \{2,4,8\}$ (see \cref{fig:maxStiffness}). We observed that while the average and maximum normalized cartographic error remain nearly identical for $s_{\textit{high}} = 4$ and $s_{\textit{high}} = 8$, the algorithm performs better at $s_{\textit{high}} = 8$ in terms of both average and maximum polygon complexity. Based on these observations, we set $s_{\textit{high}} = 8$ in our algorithm. This value achieves very small cartographic error while maintaining acceptable polygon complexity. Moreover, increasing $s_{\textit{high}}$ beyond 8 does not lead to meaningful improvements in either evaluation metric.

\begin{figure}[htb]
\centering  
\begin{minipage}[t]{.28\textwidth}
    \centering
    \includegraphics[width=0.95\textwidth]{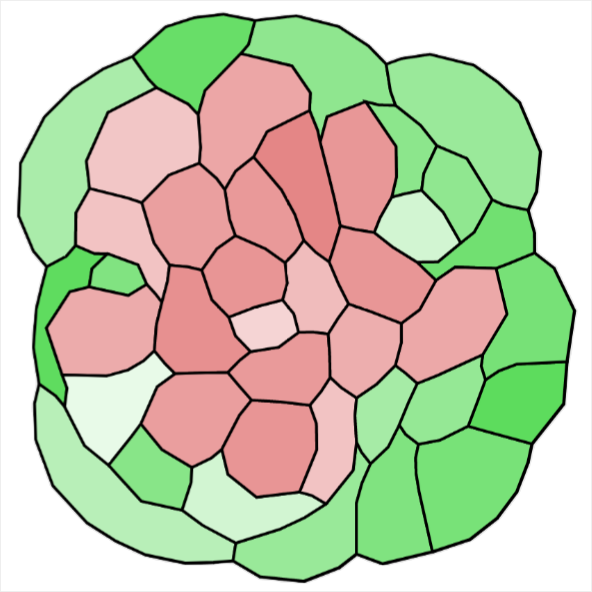}
    \subcaption{ \MS}
\end{minipage}
\begin{minipage}[t]{.28\textwidth}
    \centering
    \includegraphics[width=0.95\textwidth]{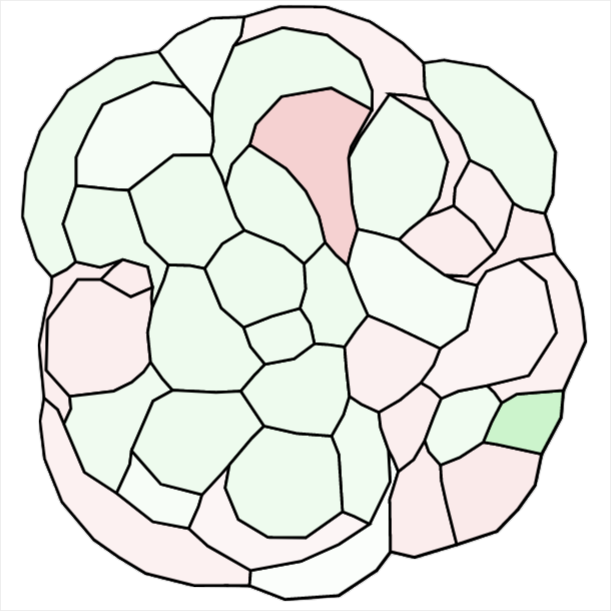}
    \subcaption{\nolinenumbers $s_{\textit{high}} = 2$}
\end{minipage}
\begin{minipage}[t]{.28\textwidth}
    \centering
    \includegraphics[width=0.95\textwidth]{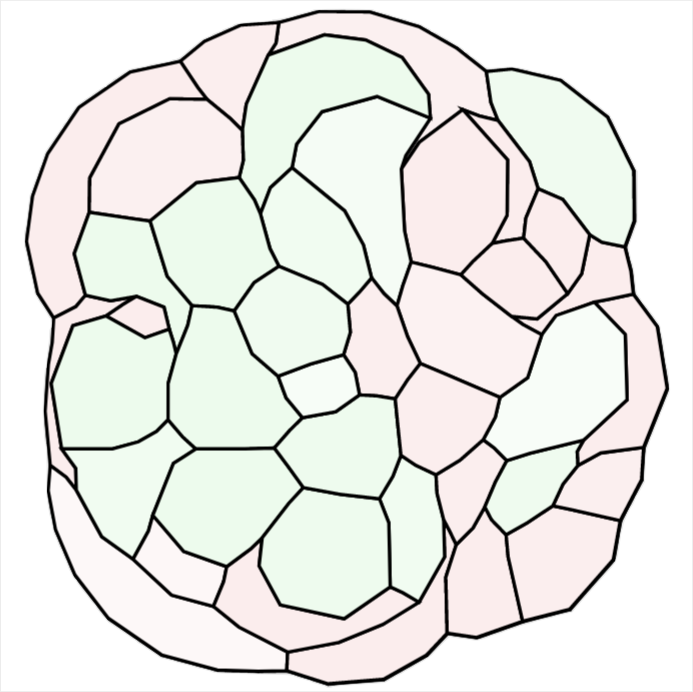}
    \subcaption{\nolinenumbers $s_{\textit{high}} = 8$}
\end{minipage}
\begin{minipage}[t]{.115\textwidth}
    \centering
    \includegraphics[width=0.95\textwidth]{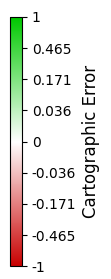}
\end{minipage}
\caption{Metaphorical maps for the same graph for \MS ~ and \NEW ~ for different values of max stiffness $s_{\textit{high}}$ }
\label{fig:heatMap}
\end{figure}

\cref{fig:heatMap} shows a ``heat-map''  style coloring of the regions of metaphorical maps of the same graph for \MS ~(\cref{fig:heatMap}.(a)) and our algorithm for  $s_{\textit{high}} \in \{2,8\}$ (\cref{fig:heatMap}.(b)-(c)), demonstrating the effect of the introduction of the stiffness coefficient.
These experiments reveal a monotonic decrease in the maximum/average cartographic error and a monotonic increase in maximum/average polygon complexity based on the value of $s_{\textit{high}}$.

\begin{figure}[htb]
	\centering  
	\begin{minipage}[t]{.48\textwidth}
		\centering
		\includegraphics[page=1]{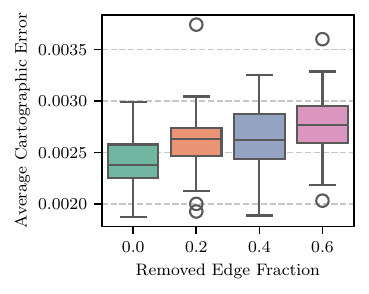}
	\end{minipage}
	\begin{minipage}[t]{.48\textwidth}
		\centering
		\includegraphics[page=1]{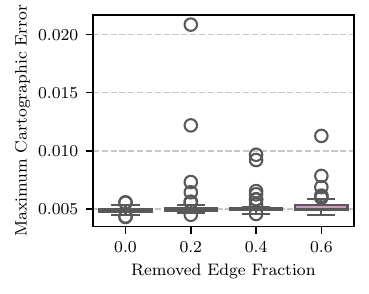}
	\end{minipage}
	\begin{minipage}[t]{.48\textwidth}
		\centering
		\includegraphics[page=1]{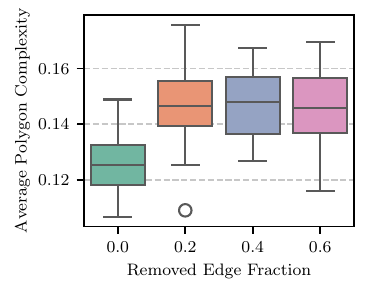}
	\end{minipage}
	\begin{minipage}[t]{.48\textwidth}
		\centering
		\includegraphics[page=1]{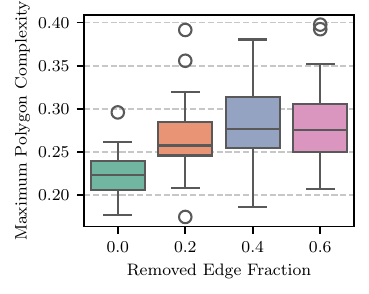}
	\end{minipage}
	\caption{Cartographic Error and Polygon Complexity for different values of the ratio $\textit{rem}$ of removed edges and by adopting function $\textit{holeWeight}()$ for assigning weights to the polygonal holes. Here, $n = 40$, $w = 5$, $\textit{nest} = 0$.}
	\label{fig:holes_statistics_removedPercentage}
\end{figure}

\begin{figure}[h!]
\centering  
\begin{minipage}[t]{.24\textwidth}
    \centering
    \includegraphics[width=0.9\linewidth]{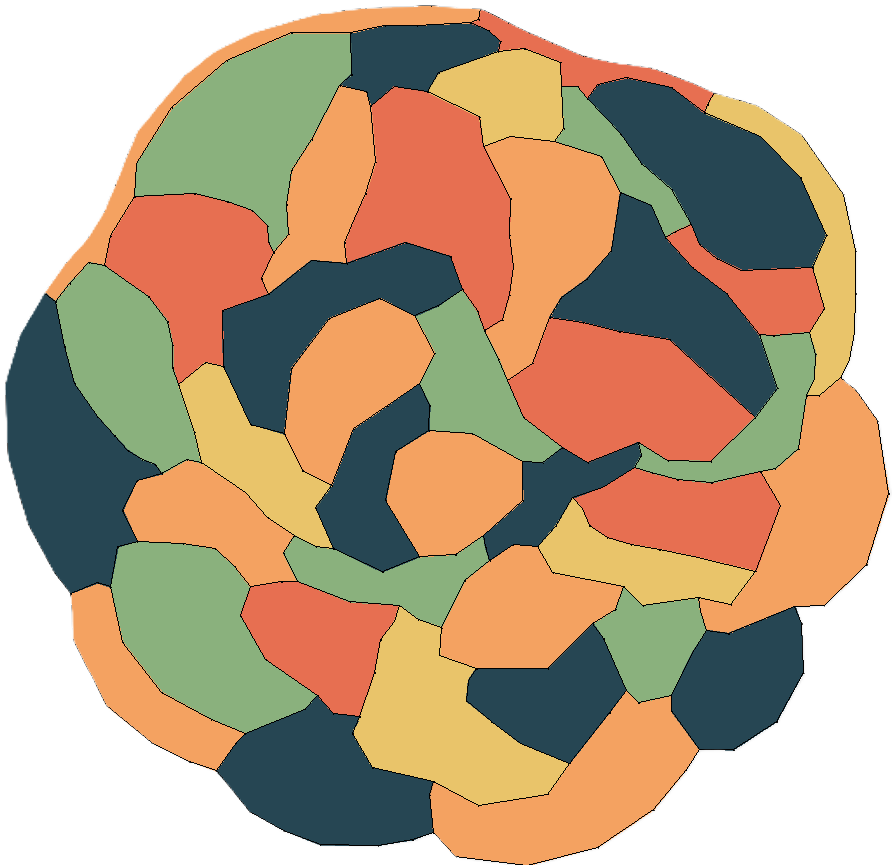}
\end{minipage}
\begin{minipage}[t]{.24\textwidth}
    \centering
    \includegraphics[width=0.9\linewidth]{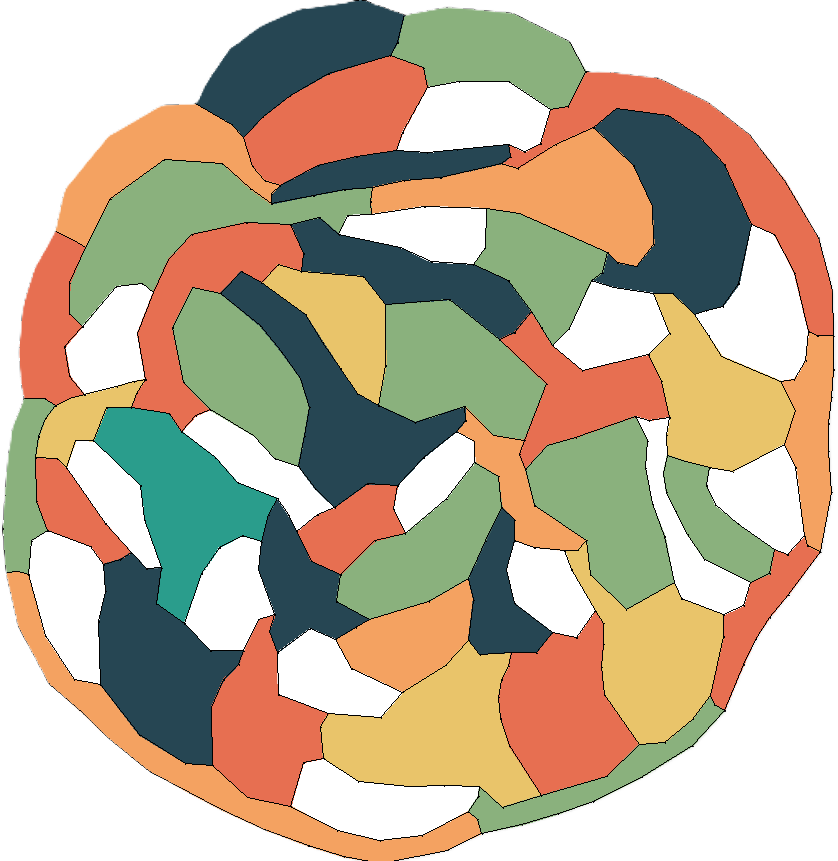}
\end{minipage}
\begin{minipage}[t]{.24\textwidth}
    \centering
    \includegraphics[width=0.9\linewidth]{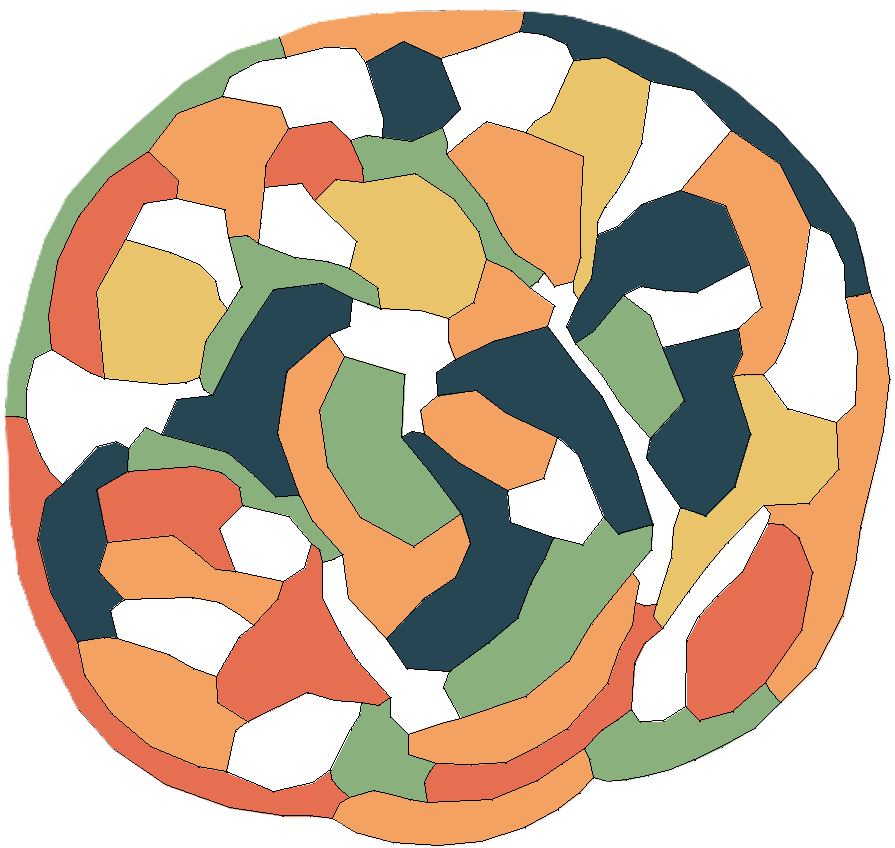}
\end{minipage}
\begin{minipage}[t]{.24\textwidth}
    \centering
    \includegraphics[width=0.9\linewidth]{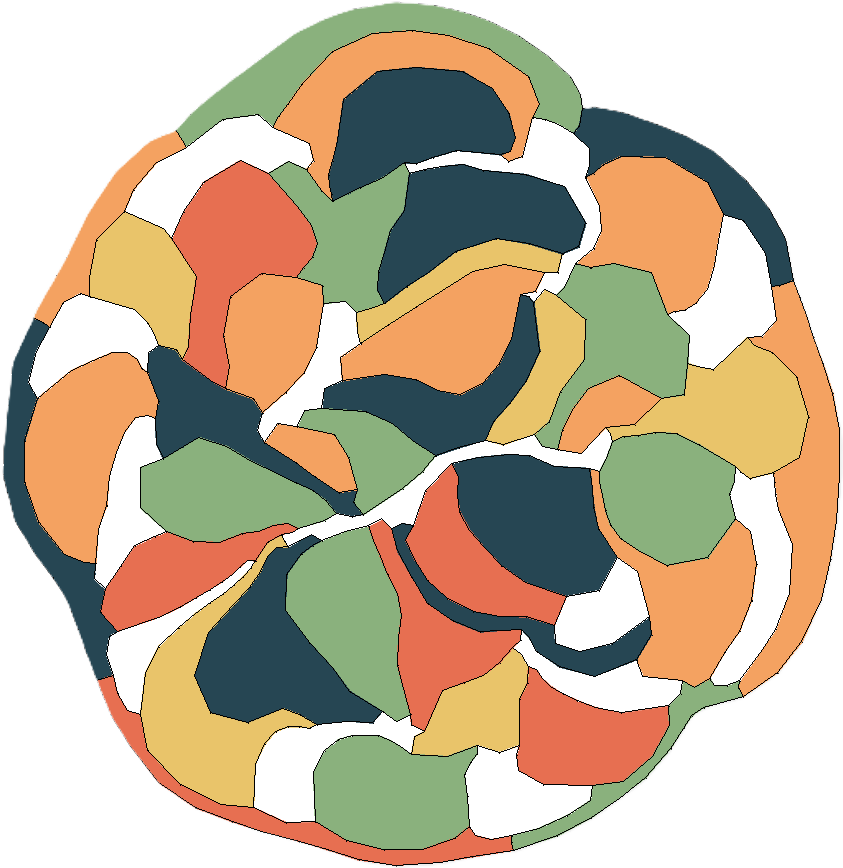}
\end{minipage}
\caption{Example for the values of $\textit{rem} \in \{0, 0.2, 0.4, 0.6\}$. }
\label{fig:rem}
\end{figure}

\subparagraph{Non-Triangulated Graphs}
We tested our algorithm for graphs with  ratio of removed edges $\textit{rem}  \in \{0, 0.2, 0.4, 0.6\}$ (\cref{fig:holes_statistics_removedPercentage} and~\ref{fig:rem}). For each value of $\mathit{rem}$, we conducted experiments on fifty graphs, each with forty nodes, a nesting ratio of zero, and a weight ratio of five. We let the algorithm run for $1200$ iterations. Recall that  hole region weights were  assigned according to  Equation~\ref{eq:holeweight}. Normalized cartographic error and polygon complexity were measured only for non-hole regions.

We observed that increasing the parameter $\textit{rem}$ (i.e., the fraction of removed edges) led to a slight increase across all four evaluation metrics (\cref{fig:holes_statistics_removedPercentage}).  This can be explained by the fact that the increasing number of missing edges creates more complex patterns of adjacency, i.e. holes behave as nodes of high degree.

We also observed that holes can have narrow passages; refer to~\cref{fig:rem}.(c)-(d), which possibly can also be explained by the relatively high number of regions adjacent to a hole. We anticipate that this effect can be mitigated by using a different hole weight function  and by re-engineering the applied forces.

\begin{figure}[h!]
\centering  
\begin{minipage}[t]{.24\textwidth}
    \centering
    \includegraphics[width=0.9\linewidth]{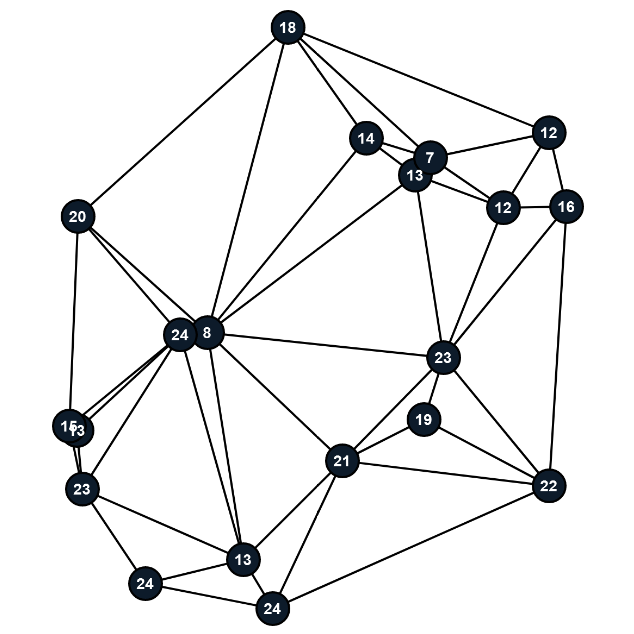}
    \subcaption{Initial Graph ($n=20$).}
\end{minipage}
\begin{minipage}[t]{.24\textwidth}
    \centering
    \includegraphics[width=0.9\linewidth]{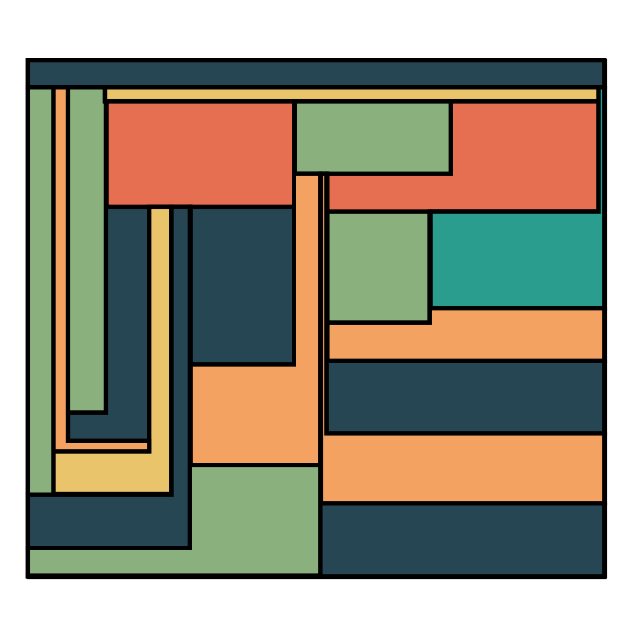}
    \subcaption{\nolinenumbers Octagonal map.}
\end{minipage}
\begin{minipage}[t]{.24\textwidth}
    \centering
    \includegraphics[width=0.9\linewidth]{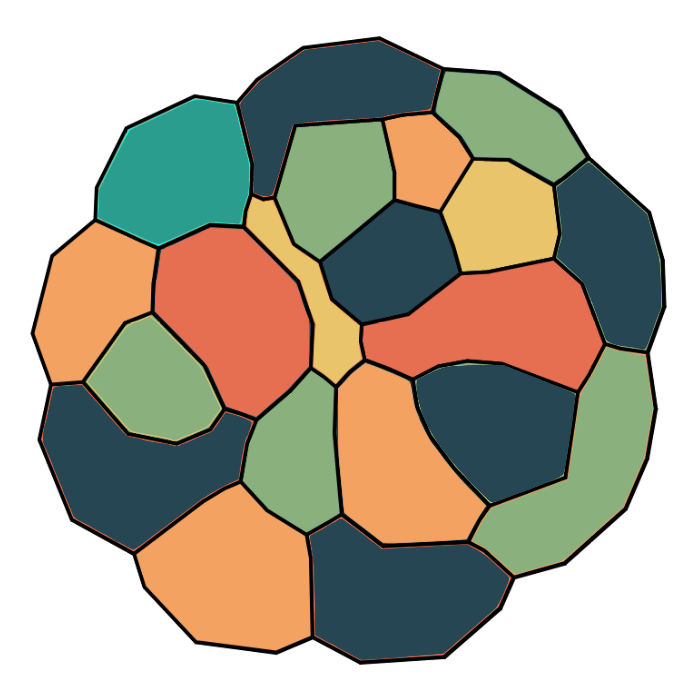}
    \subcaption{\nolinenumbers Output based on graph's dual.}
\end{minipage}
\begin{minipage}[t]{.24\textwidth}
    \centering
    \includegraphics[width=0.9\linewidth]{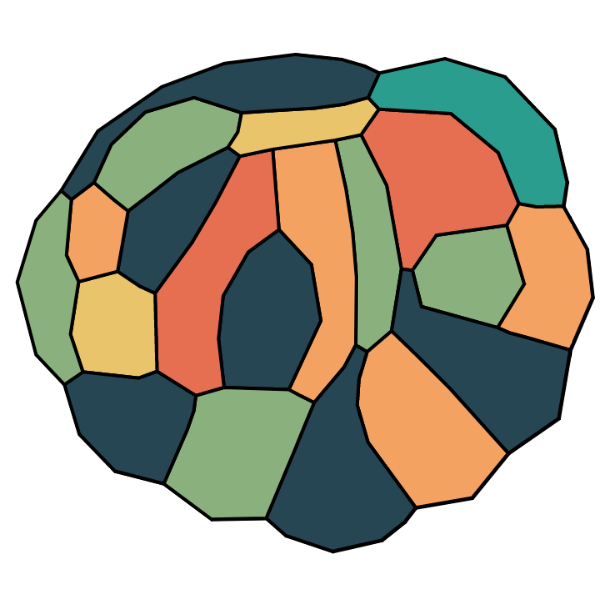}
    \subcaption{\nolinenumbers Output based on octagonal map.}
\end{minipage}
\begin{minipage}[t]{.24\textwidth}
    \centering
    \includegraphics[width=0.9\linewidth]{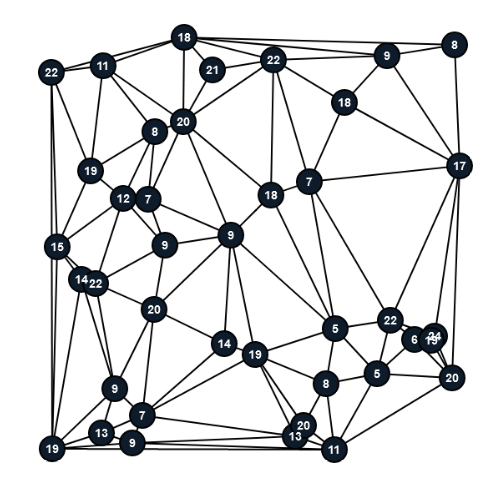}
    \subcaption{Initial Graph ($n=40$).}
\end{minipage}
\begin{minipage}[t]{.24\textwidth}
    \centering
    \includegraphics[width=0.9\linewidth]{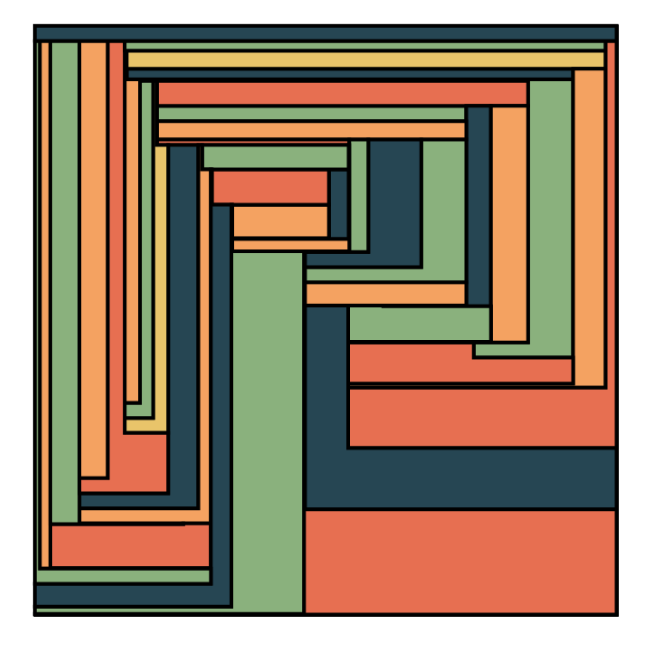}
    \subcaption{\nolinenumbers Octagonal map.}
\end{minipage}
\begin{minipage}[t]{.24\textwidth}
    \centering
   \includegraphics[width=0.9\linewidth]{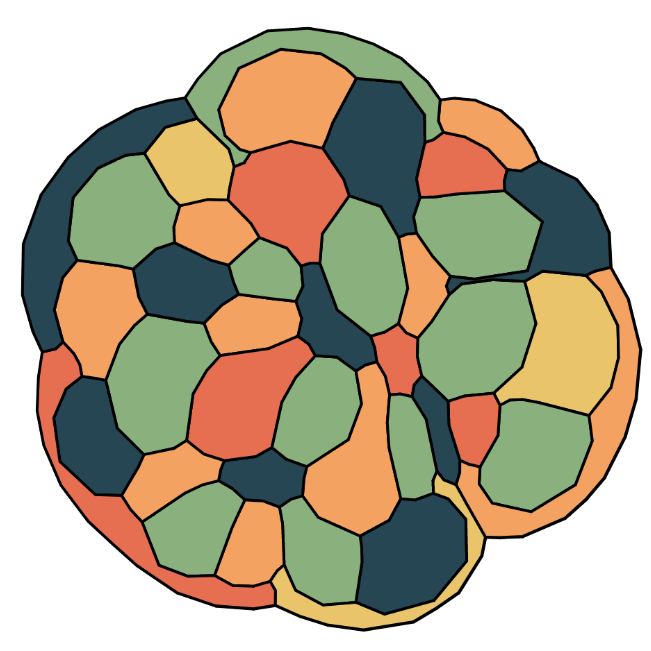}
    \subcaption{\nolinenumbers Output based on graph's dual.}
\end{minipage}
\begin{minipage}[t]{.24\textwidth}
    \centering
    \includegraphics[width=0.9\linewidth]{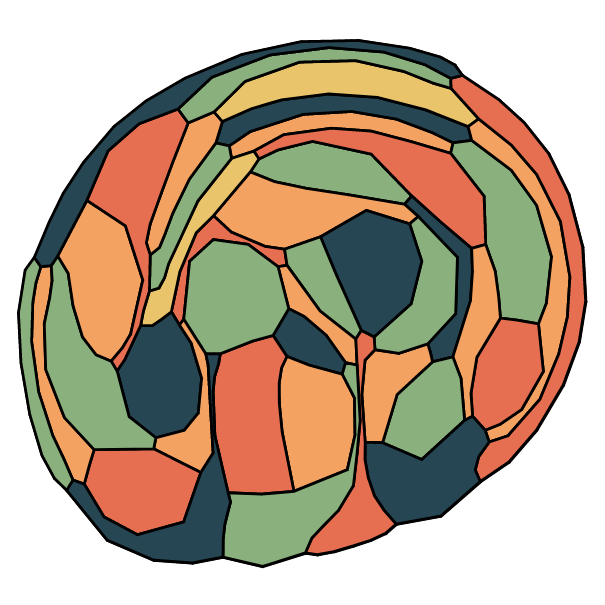}
    \subcaption{\nolinenumbers Output based on octagonal map.}
\end{minipage}
\caption{Metaphorical maps produced by our algorithm when the initial map is based on the graph's dual transformation and on an octagonal map.  (a)-(d): 20 vertex graph. (e)-(h): 40 vertex graph.}
\label{fig:example_Alam_20}
\end{figure}

\subparagraph{The effect of the initial drawing}
Given that we managed to, effectively, get the average cartographic error down to zero, we did not study the effect of the initial metaphorical map used by our algorithm on the cartographic error. Instead, we focused on the polygon complexity and we  examined the effect of having an initial map with good average cartographic error. As such good initial maps we considered the maps produced by the algorithm in~\cite{AlamBFKKT2013} (referred to as \emph{octagonal map})   which produces rectangular maps with at most 8 corners and of average cartographic error close to zero. 
We observed that, when using octagonal maps as initial layout, no consistent improvement was observed while the produced maps appeared to have long and skinny regions. Moreover, when the number of vertices becomes larger the metaphorical maps become harder to read due to their long, skinny, and circular shaped regions. See \cref{fig:example_Alam_20}.(a-d) and \cref{fig:example_Alam_20}.(e-h) for sample metaphorical maps of 20 and 40 regions, respectively. Finally, it is noted that our algorithm also required a larger number of iterations (about double) when initialized with such maps in order to converge to its final metaphorical map solution. 

\section{Conclusion}
\label{sec:future_work}
In this work we have proposed an improvement of \MS~\cite{MchedSchnorr22}  that produces metaphorical maps of vertex-weighted graphs and presented a detailed experimental evaluation of the newly proposed algorithm. Our goal was to close the gap between the metaphorical maps with perfect cartographic accuracy~\cite{alam2013linear,kleist2018drawing,thomassen1992plane} that despite using a few corners per polygonal region can look quite complex and the simple-looking metaphorical maps produced by \MS~ with cartographic error up to 30\%. 
Our experiments showed that we can achieve cartographic error close to zero, by paying a small price for the polygon complexity. Our algorithm achieves this improvement by using the notion of region stiffness that adapts as the force-directed simulation runs. We have seen that the maximum value of this stiffness determines the interplay between the cartographic error and polygon complexity. 

To quantify the complexity of the maps, we applied the quality metric introduced in~\cite{BrinkhoffKSB95}, who presented examples of geographic maps showing that this metric corresponds to the human intuition of map complexity. Our experiments with metaphorical maps do not contradict this intuition. However, we believe that 
whether this function is indeed appropriate for qualifying the human intuition of the metaphorical maps complexity has to be investigated in a user study.  

We conjecture that the performance of the algorithm for non-triangulated graphs can be further improved by considering different hole weight functions  and  by engineering the forces to avoid narrow passages. Finally, it is also interesting to compare the quality of the maps for non-triangulated graphs with and without holes. We defer these investigations to the extended version of this work.

\newpage
\bibliographystyle{plainurl}
\bibliography{cartogram_references.bib}

\begin{thebibliography}{10}

\bibitem{thesisAlam15}
Muhammad~Jawaherul Alam.
\newblock {\em Contact representations of graphs in 2D and 3D}.
\newblock Doctoral thesis, The University of Arizona, 2015.
\newblock URL:
  \url{https://www.proquest.com/openview/5bf2b2dc7f21bf9f8759bdc89632b50b/1?pq-origsite=gscholar&cbl=18750}.

\bibitem{alam2013linear}
Muhammad~Jawaherul Alam, Therese Biedl, Stefan Felsner, Andreas Gerasch,
  Michael Kaufmann, and Stephen~G. Kobourov.
\newblock Linear-time algorithms for hole-free rectilinear proportional contact
  graph representations.
\newblock {\em Algorithmica}, 67(1):3--22, 2013.

\bibitem{AlamBFKKT2013}
Muhammad~Jawaherul Alam, Therese Biedl, Stefan Felsner, Michael Kaufmann,
  Stephen~G. Kobourov, and Torsten Ueckerdt.
\newblock Computing cartograms with optimal complexity.
\newblock {\em Discrete \& Computational Geometry}, 50(3):784--810, 2013.
\newblock \href {https://doi.org/10.1007/s00454-013-9521-1}
  {\path{doi:10.1007/s00454-013-9521-1}}.

\bibitem{Alam2015}
Muhammad~Jawaherul Alam, Stephen~G. Kobourov, and Sankar Veeramoni.
\newblock Quantitative measures for cartogram generation techniques.
\newblock {\em Comput. Graph. Forum}, 34(3):351--360, 2015.
\newblock \href {https://doi.org/10.1111/cgf.12647}
  {\path{doi:10.1111/cgf.12647}}.

\bibitem{BernE1995}
Marshall Bern and David Eppstein.
\newblock {\em Mesh Generation and Optimal Triangulation}, pages 47--123.
\newblock 1995.
\newblock \href {https://doi.org/10.1142/9789812831699_0003}
  {\path{doi:10.1142/9789812831699_0003}}.

\bibitem{Biuk-AghaiPP17}
Robert~P. Biuk{-}Aghai, Patrick~Cheong{-}Iao Pang, and Bin Pang.
\newblock Map-like visualisations vs. treemaps: an experimental comparison.
\newblock In Robert~P. Biuk{-}Aghai, Jie Li, and Shigeo Takahashi, editors,
  {\em Proceedings of the 10th International Symposium on Visual Information
  Communication and Interaction, {VINCI} 2017, Bangkok, Thailand, August 14-16,
  2017}, pages 113--120. {ACM}, 2017.
\newblock \href {https://doi.org/10.1145/3105971.3105976}
  {\path{doi:10.1145/3105971.3105976}}.

\bibitem{BortinsDC}
I.~Bortins, S.~Demers, and K.~Clarke.
\newblock Cartogram types, 2002.

\bibitem{BrinkhoffKSB95}
Thomas Brinkhoff, Hans{-}Peter Kriegel, Ralf Schneider, and A.~Braun.
\newblock Measuring the complexity of polygonal objects.
\newblock In Patrick Bergougnoux, Kia Makki, and Niki Pissinou, editors, {\em
  Proceedings of the 3rd {ACM} International Workshop on Advances in Geographic
  Information Systems, Baltimore, Maryland, USA, December 1-2, 1995, in
  conjunction with {CIKM} 1995}, page 109. {ACM}, 1995.

\bibitem{CanoBCPSS15}
Rafael~G. Cano, Kevin Buchin, Thom Castermans, Astrid Pieterse, Willem Sonke,
  and Bettina Speckmann.
\newblock Mosaic drawings and cartograms.
\newblock {\em Comput. Graph. Forum}, 34(3):361--370, 2015.
\newblock \href {https://doi.org/10.1111/cgf.12648}
  {\path{doi:10.1111/cgf.12648}}.

\bibitem{David96:seeing}
Prabu David.
\newblock Seeing is believing: Comparative performance of the pie and the bar.
\newblock {\em Newspaper Research Journal}, 17(1-2):89--104, 1996.
\newblock \href {https://doi.org/10.1177/073953299601700109}
  {\path{doi:10.1177/073953299601700109}}.

\bibitem{BergMS09}
Mark de~Berg, Elena Mumford, and Bettina Speckmann.
\newblock On rectilinear duals for vertex-weighted plane graphs.
\newblock {\em Discret. Math.}, 309(7):1794--1812, 2009.
\newblock \href {https://doi.org/10.1016/j.disc.2007.12.087}
  {\path{doi:10.1016/j.disc.2007.12.087}}.

\bibitem{Dorling}
Danny Dorling.
\newblock Area cartograms: their use and creation.
\newblock {\em University of East Anglia: Environmental Publications}, 59,
  1996.

\bibitem{HograferHS20}
Marius Hogr{\"{a}}fer, Magnus Heitzler, and Hans{-}J{\"{o}}rg Schulz.
\newblock The state of the art in map-like visualization.
\newblock {\em Comput. Graph. Forum}, 39(3):647--674, 2020.
\newblock \href {https://doi.org/10.1111/cgf.14031}
  {\path{doi:10.1111/cgf.14031}}.

\bibitem{Hu05}
Yifan Hu.
\newblock Efficient and high quality force-directed graph drawing.
\newblock {\em Mathematica Journal}, 10:37--71, 01 2005.

\bibitem{kleist2018drawing}
Linda Kleist.
\newblock Drawing planar graphs with prescribed face areas.
\newblock {\em J. Comput. Geom.}, 9(1):290--311, 2018.
\newblock \href {https://doi.org/10.20382/jocg.v9i1a9}
  {\path{doi:10.20382/jocg.v9i1a9}}.

\bibitem{MchedSchnorr22}
Tamara Mchedlidze and Christian Schnorr.
\newblock {Metaphoric Maps for Dynamic Vertex-weighted Graphs}.
\newblock In Marco Agus, Wolfgang Aigner, and Thomas Hoellt, editors, {\em
  EuroVis 2022 - Short Papers}. The Eurographics Association, 2022.
\newblock \href {https://doi.org/10.2312/evs.20221090}
  {\path{doi:10.2312/evs.20221090}}.

\bibitem{nusrat2016state}
Sabrina Nusrat and Stephen~G. Kobourov.
\newblock The state of the art in cartograms.
\newblock {\em Comput. Graph. Forum}, 35(3):619--642, 2016.
\newblock \href {https://doi.org/10.1111/cgf.12932}
  {\path{doi:10.1111/cgf.12932}}.

\bibitem{Owen1998ASO}
Steven~J. Owen.
\newblock A survey of unstructured mesh generation technology.
\newblock In {\em 7th International Meshing Roundtable}, 1998.

\bibitem{SaketSK16}
Bahador Saket, Carlos Scheidegger, and Stephen~G. Kobourov.
\newblock Comparing node-link and node-link-group visualizations from an
  enjoyment perspective.
\newblock {\em Comput. Graph. Forum}, 35(3):41--50, 2016.
\newblock \href {https://doi.org/10.1111/cgf.12880}
  {\path{doi:10.1111/cgf.12880}}.

\bibitem{Schnorr2020Thesis}
C.~Schnorr.
\newblock Visualizing dynamic clustered data using areaproportional maps.
\newblock Master's thesis, Karlsruhe Institute of Technology, 2020.
\newblock URL: \url{https://github.com/jenox/Master-Thesis}.

\bibitem{SimonettoAAB2011impred}
Paolo Simonetto, Daniel Archambault, David Auber, and Romain Bourqui.
\newblock Impred: An improved force-directed algorithm that prevents nodes from
  crossing edges.
\newblock {\em Comput. Graph. Forum}, 30(3):1071--1080, 2011.
\newblock \href {https://doi.org/10.1111/j.1467-8659.2011.01956.x}
  {\path{doi:10.1111/j.1467-8659.2011.01956.x}}.

\bibitem{2013gd}
Roberto Tamassia, editor.
\newblock {\em Handbook on Graph Drawing and Visualization}.
\newblock Chapman and Hall/CRC, 2013.

\bibitem{thomassen1992plane}
Carsten Thomassen.
\newblock Plane cubic graphs with prescribed face areas.
\newblock {\em Comb. Probab. Comput.}, 1:371--381, 1992.
\newblock \href {https://doi.org/10.1017/S0963548300000407}
  {\path{doi:10.1017/S0963548300000407}}.

\bibitem{tobler2004thirty}
Waldo Tobler.
\newblock Thirty five years of computer cartograms.
\newblock {\em An. of the Assoc. of American Geographers}, 94(1):58--73, 2004.
\newblock \href {https://doi.org/10.1111/j.1467-8306.2004.09401004.x}
  {\path{doi:10.1111/j.1467-8306.2004.09401004.x}}.

\bibitem{Tutte1963}
William~Thomas Tutte.
\newblock How to draw a graph.
\newblock {\em Proceedings of the London Mathematical Society}, 3(1):743--767,
  1963.

\bibitem{WuTC2008}
Han-Ming Wu, ShengLi Tzeng, and Chun-houh Chen.
\newblock {\em Matrix Visualization}, pages 681--708.
\newblock Springer Berlin Heidelberg, Berlin, Heidelberg, 2008.
\newblock \href {https://doi.org/10.1007/978-3-540-33037-0_26}
  {\path{doi:10.1007/978-3-540-33037-0_26}}.

\bibitem{yFiles2025}
{yWorks GmbH}.
\newblock yfiles for html.
\newblock \url{https://www.yworks.com/products/yfiles-for-html}, 2025.
\newblock Software library.

\end{thebibliography}

\newpage
\begin{appendix}

{\huge Appendix}

\section{Polygon Complexity}
\label{app:complexity}
Brinkhoff, Kriegel, Schneider, and Braun~\cite{BrinkhoffKSB95}  defined the complexity of a polygon $P$ as a function of three quantities:  (a) the \emph{frequency  of $P$'s vibration}, denoted by $\text{freq}(P)$,  (b) the \emph{amplitude} of $P$'s vibration, denoted by $\text{ampl}(P)$, and (c) $P$'s \emph{convexity}, denoted by $\text{conv}(P)$. 
Let $P$ be an $n$-vertex polygon, $n\geq4$,  and let $\text{hull}(P)$, $A(P)$,  $\text{circ}(P)$ and $\text{encircle}(P)$  denote $P$'s convex hull,  area,  boundary length, and  smallest enclosing circle, respectively. 
Let  ${L}(P)$ be the number of $P$'s internal concave angles,
and let ${L}'(P)  = {L}(P)/(n-3)$.

\noindent The \emph{frequency of $P$'s vibration}  is defined as 
\begin{equation*}
\text{freq}(P) = 1
+ 16 \cdot ({L}'(P) - 0.5)^4
- 8 \cdot ({L}'(P) - 0.5)^2
\end{equation*}

\noindent  Function $\text{freq}(P)$ takes values in the range $\lbrack0,1\rbrack$, is zero for a convex polygon, and takes its   maximum value when half of the internal angles are concave.

\noindent The  \emph{amplitude of $P$'s vibration} is defines as

\begin{equation*}
\text{ampl}(P) =
\frac{\text{circ}(P) - \text{circ}(\text{hull}(P))}{\text{circ}(P)}
\end{equation*}

\noindent Function $\text{ampl}(P)$ also takes values in the range $ \lbrack0,1\rbrack$; it is  0 for a convex polygon and tends to 1 when the boundary of the polygon is getting longer compared to the length of the convex hull.

To measure the \emph{convexity of a polygon},  Mchedlidze and Schnorr~\cite{MchedSchnorr22} slightly modified the measure proposed in~\cite{BrinkhoffKSB95}  in order to differentiate between lengthy and fat convex polygons. 
They measured the fraction of area that polygon $P$ does not cover in its smallest enclosing circle. In order to keep the metric in the unit interval, they actually compare $P$'s area to the  area of a regular $n$-gon having $P$'s enclosing circle as its circumcircle. The convexity of polygon $P$ was defines as:

\begin{equation*}
\text{conv}(P) =
1 - \frac{A(P)}{A(\text{encircle}(P)) \cdot \sin\left(\frac{360^\circ}{n}\right) \cdot \frac{n}{2\pi}}    
\end{equation*}

\noindent Based on  quantities $\text{freq}(P)$, $\text{ampl}(P)$ and $\text{conv}(P)$, Brinkhoff, Kriegel, Schneider, and Braun~\cite{BrinkhoffKSB95}  defined $P$'s complexity, denoted by $\text{compl}(P)$ and taking values in  $\lbrack0,1\rbrack$, as

\begin{equation}
\text{compl}(P) =
0.8 \cdot \text{ampl}(P) \cdot \text{freq}(P) + 0.2 \cdot \text{conv}(P)
\end{equation}

 \newpage
 \section{Further details of the \MS}
 \label{app:forces}
 The following forces which are both employed in the \MS~and in our algorithm are designed towards obtaining metaphorical maps with low polygon complexity. 
 
 The \emph{vertex-vertex repulsion force}  $\cvector{F}(u,v)$ is applied at the vertex $u$ and aims to prevent the vertices from clumping together. It is defined  as $\cvector{F}(u,v) = 25  \frac{1}{\|\cvector{u}[v]\|^2} \nvector{u}[v]$\footnote{Let $\cvector{q}$ be a vector. Then, by $\nvector{q}$ we denoted the normalized vector of $\cvector{q}$, that is, the free unit vector that has the same direction as $\cvector{q}$.}.

The purpose of the \emph{vertex-edge repulsion force} $\cvector{F}(v,e)$ is to prevent the creation of polygons with narrow passages. Given an edge $e$ and non-incident vertex $v$, let $\nvector{r}$ denote the unit vector perpendicular to edge $e$ that is directed form $v$ toward the line containing edge $e$. Further, let $x \in e$ be the point on edge $e$ that is closest to vertex $v$. The vertex-edge repulsion force is defined as  $\cvector{F}(v,e) = 10  \frac{1}{\|\cvector{x v}\|^2}  (\nvector{r} \cdot \nvector{x v})  \nvector{x v}$ and is exerted on $v$.

For efficiency, the $\cvector{F}(v,e)$ as well as the $\cvector{F}(u,v)$ forces are computed only for elements that belong to the same face of the metaphorical map. 

The \emph{angular-resolution force} also aims to eliminate narrow passages by trying to evenly distribute the angles around a vertex.  
Consider a  vertex $v$ and let $u$ and $w$ be two of its consecutive neighbors in clockwise order. Let $\alpha_{uvw}$
denotes the angle $\angle{uvw}$ and $\nvector{b}_{uvw}$ be the normalized bisector of  $\angle{uvw}$. We define the angular resolution   force $\cvector{F}(u,v,w)$ at vertex $v$ as 
$ \cvector{F}(u,v,w) =	\frac{1}{2}  \frac{\frac{360^\circ}{\deg(v)} - \alpha_{uvw}}{\alpha_{uvw}}	 \nvector{b}_{uvw}$.

The purpose of the \emph{air-pressure force} is to impel  the area of each of the map's faces to assume its desired value (or to get as close to it as possible). Similar to ~\cite{AlamBFKKT2013}, Mchedlidze and Schnorr treated the  map's polygonal regions as volumes of some amount of air equal to the respective region's weight. Let $F$ be the set of regions of a metaphorical map. The normalized pressure $P(g)$ in an internal polygonal region $g$  was defined as  $P(g) = \frac{w(g)}{A(g)}  \frac{\sum_{f \in F}{A(f)}}{\sum_{f \in F}{w(f)}}$ while for the outer region $g_{out}$ they set $P(g_{out}) =  1$.  
Thus,  the \emph{air-pressure force} on edge $e$ of region $g$ was defined as  $\cvector{F}(e,g) =3  P(g)\frac{\ell(e)}{\text{circ}(g)}\nvector{r}$,  
where $\nvector{r}$ is a unit vector perpendicular to $e$ directed towards outside of $g$. Force $\cvector{F}(e,g)$ was applied to both endpoints of edge $e$.

\begin{figure}
\centering  
\begin{minipage}[t]{.48\textwidth}
    \centering
    \includegraphics[width=0.5\linewidth]{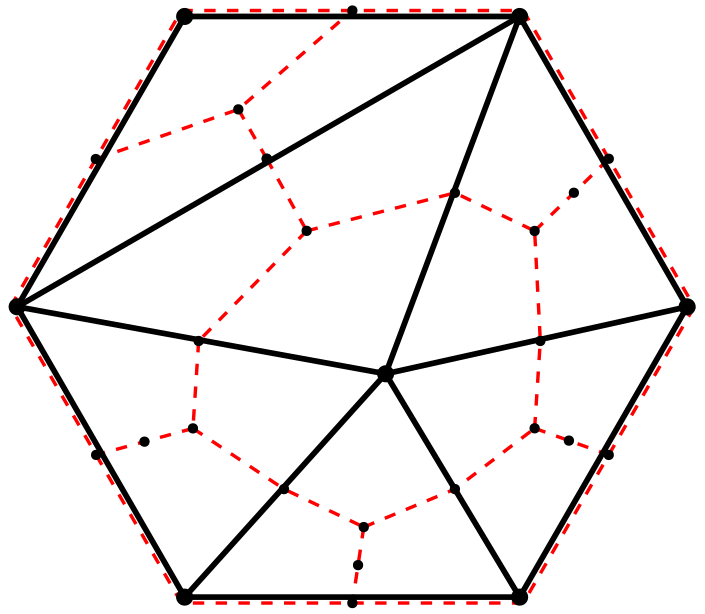}
\end{minipage}
\begin{minipage}[t]{.48\textwidth}
    \centering
    \includegraphics[width=0.5\linewidth]{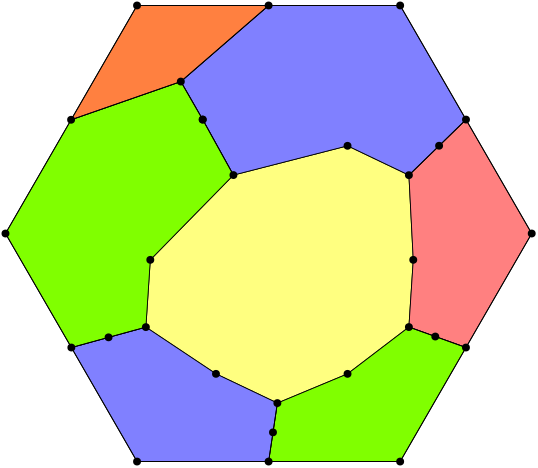}
\end{minipage}
\caption{Transformation of a graph (solid lines) to the metaphorical map (dashed lines) based on the notion of dual graph and the resulting metaphorical map.}
\label{fig:transformation}
\end{figure}

The initial layout of the map is computed based on the notion of the dual; Refer to \cref{fig:transformation}. The algorithm starts with a planar drawing of an internally triangulated graph $G$.  The dual vertices of the inner faces are placed in the barycenters of the triangles representing those faces. The dual
edges of are drawn as polylines consisting of two segments meeting at a bend
vertex, which lies in the middle of the corresponding primal edge.

\section{Setting %
$\textit{step}$, $\textit{iter}$ and $s_\textit{high}$}
\label{app:parameters}

We determined the values of the parameters $\textit{step}$, $\textit{iter}$ and $s_\textit{high}$ by running our algorithm on a suite of fifty 20--node random graphs with weight ratio five and the nesting ratio zero. In each experiment we used

\begin{figure}[htb]
\centering  
\begin{minipage}[t]{0.9\linewidth}
    \centering
    \includegraphics[page=1]{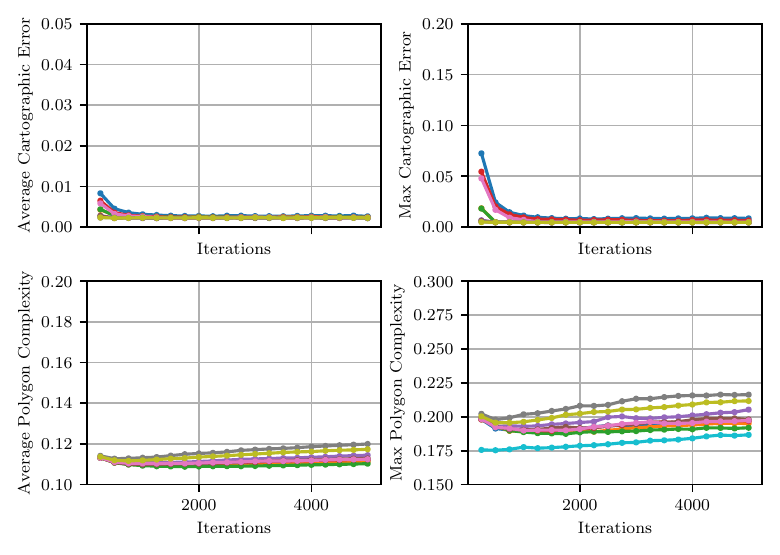}
\end{minipage}
\caption{Line plots for MS and for every ($\textit{step} \times s_{\textit{high}} $), $\textit{step} \in \{0.01, 0.02, 0.04 \}, s_{\textit{high}} \in \{2, 4, 8\}$, $\textit{iter} = 5000$, $n = 20$, $w = 5$, $\textit{nest}=0$.}
\label{fig:additive_singleGraph}
\end{figure}

\begin{itemize}
	\item $\textit{step} \in \{0.01, 0.02, 0.04\}$.  We selected to study small additive values since we wanted to avoid a quick increase of the stiffness coefficient which could potentially lead to having regions alternating between over-pressured and under-pressured status.
	\item $s_{\textit{high}} \in \{2, 4, 8\}$. Given that our algorithm is identical to the \MS~ for stiffness $s_{\textit{high}}=1$ (without applying the corrective weight coefficients on the pressure forces), we investigated the effect of the maximum stiffness  $s_{\textit{high}}$ on our evaluation metrics.
	\item $5000$ iterations. We observed that the average cartographic error converges very quickly (after about 500 iterations) to a value near zero and that no major changes occur afterwards. \cref{fig:additive_singleGraph}
	shows \footnote{In all our figure legends, $n$ denotes the number of vertices of the graph, $w$ denotes the ratio of maximum to minimum possible vertex weight,  $\textit{iter}$ denotes the number of iterations in our algorithm and $\textit{nest}$ denotes the graph nesting ratio (see \cref{sec:testBed}).} 
	the deviation of all the quality metrics of our study. 
\end{itemize} 

\begin{figure}[th]
\centering  
\begin{minipage}[t]{0.9\linewidth}
    \centering
    \includegraphics[page=1]{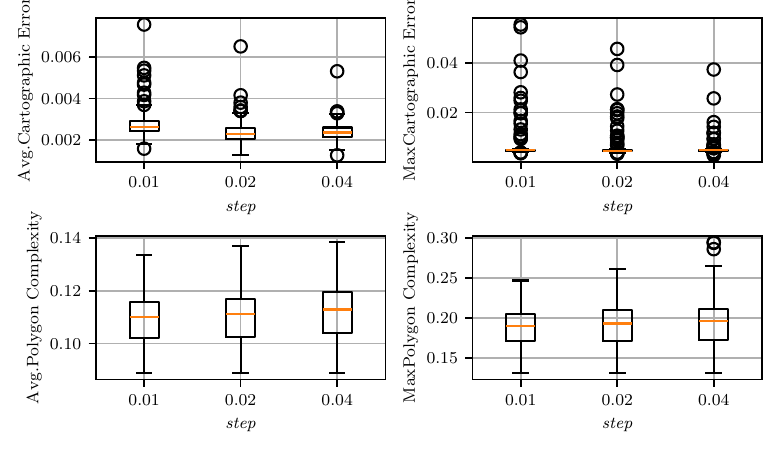}
\end{minipage}
\caption{Box plots of evaluation metrics on 50 %
graphs for  $\textit{step} \in \{0.01, 0.02, 0.04 \}$, $ s_{\textit{high}} \in \{2, 4, 8\}$,
$\textit{iter} = 5000$, $n = 20$, $w = 5$, $\textit{nest}=0$.}
\label{fig:additive_20graphs}
\end{figure}

\begin{figure}
	\centering  
	\begin{minipage}[t]{0.9\linewidth}
		\centering
		\includegraphics[page=1]{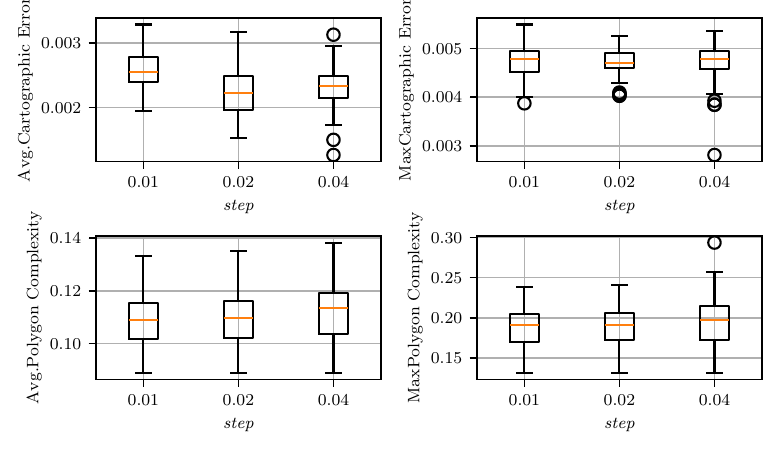}
	\end{minipage}
	\caption{Box plots of evaluation metrics on 50 %
		,  $ s_{\textit{high}} =8$,
		graphs for  $\textit{step} \in \{0.01, 0.02, 0.04 \}$, 
		$\textit{iter} = 5000$, $n = 20$, $w = 5$, $\textit{nest}=0$.}
	\label{fig:additive_20graphs2}
\end{figure}

In order to decide on the value of constant $\textit{step}$ we ran experiments for $\textit{step} \in \{0.01, 0.02, 0.04 \}$ on 50 random graphs and computed all evaluation metrics. \cref{fig:additive_20graphs} shows the results of our experiments in the form of box plots when $s_{\textit{high}} \in \{2, 4, 8\}$ and \cref{fig:additive_20graphs2} shows the results of our experiments when $s_{\textit{high}}=8$. 
\cref{fig:additive_20graphs} suggest that for $\textit{step} \in \{0.02, 0.04 \}$ we get slightly smaller  average cartographic error. However, realize that the gain appears to be almost insignificant; less than 0.1\%.  The same behaviour is observed in for the maximum cartographic error. The  three $\textit{step}$ values appear to lead to almost identical results when compared against the average polygon complexity metric. By realizing that any of the studied values lead to almost identical and high quality results, we decided to set $\textit{step}=0.02$. 

Having fixed the $\textit{step}$ constant, we next determined an appropriate number of iterations, $\textit{iter}$, that balances convergence quality with overall runtime. To this end, we measured, for each evaluation metric, the iteration at which its value changes by less than a threshold of $0.005$. Table~\ref{tab:plateau_step} reports these plateau points for $\textit{step} = 0.02$ across four values of $s_{\textit{high}}$. 
We observe that all evaluation metrics stabilize by approximately $1.000$ iterations on our 20--vertex test graphs, with the exception of the maximum cartographic error for the \MS, where small fluctuations of less than $0.008$ were still present at $iter = 1.000$. Nevertheless, such variations are minor and do not significantly affect the overall map quality. We therefore conclude that setting $iter = 1.000$ provides a good trade-off between convergence and computational efficiency for graphs of this size.
However, in order to also account for the size of the input graph, assuming that larger graphs take longer time to converge to a good metaphorical map, we decided to set the number of iteration to $\textit{iter}= 800 + 10n$, a value consistent with our decision to set $\textit{iter}= 1.000$ for 20-vertex graphs. 

\begin{table}
	\centering
	\begin{tabular}{c  p{0.2\linewidth} p{0.2\linewidth} p{0.2\linewidth} p{0.2\linewidth}}
		\hline
		$s_{\textit{high}}$ & \textbf{Avg Cart. Error} & \textbf{Max Cart. Error} & \textbf{Avg Polygon Complexity} & \textbf{Max Polygon Complexity} \\
		\hline
		MS & $\sim$1000 & $\sim$1250 & $\sim$500 & $\sim$500 \\
		2.0 & $\sim$500 & $\sim$1000 & $\sim$500 & $\sim$500 \\
		4.0 & $\sim$500 & $\sim$500 & $\sim$500 & $\sim$750 \\
		8.0 & $\sim$500 & $\sim$500 & $\sim$500 & $\sim$750 \\
		\hline
	\end{tabular}
	\caption{Approximate iteration counts at which each metric plateaus for $step = 0.02$ and a threshold of $0.005$.}
	\label{tab:plateau_step}
\end{table}

Finally, the constant $s_\textit{high}$ which upper bounds  the stiffness coefficient was set to $s_\textit{high}=8$. We defer the justification of this decision until \cref{sec:carError_PolCompl_tradeoff} where we present a trade-off between the average cartographic error and the average polygon complexity which appears to be controlled by $s_\textit{high}$. However, we notice that no improvement in the evaluation metrics occures for $s_\textit{high} >8$.

\section{Additional evaluation considerations}
\label{app:evaluationplus}

\begin{figure}[h]
	\centering  
	\begin{minipage}[t]{.9\linewidth}
		\centering
		\includegraphics[page=1]{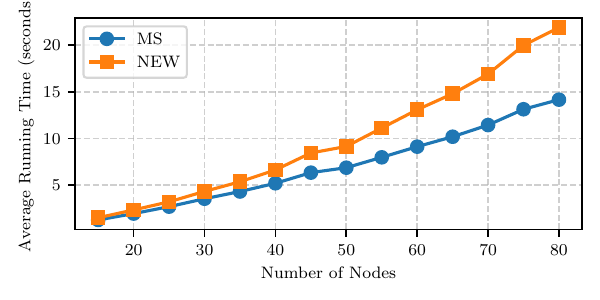}
	\end{minipage}
	\caption{The time required (in seconds) for the execution of the algorithms for $\textit{iter}=800 + 10n$ iterations. }
	\label{fig:timecomplexity}
\end{figure}

\subparagraph{Time performance}
We compared the running time of the \MS~ and \NEW~ across graphs of varying sizes. For each number of nodes $n$, we computed the average execution time over all input instances with that size. As shown in~\cref{fig:timecomplexity}, the running time of both algorithms increases with the graph size, as expected. While \NEW~ consistently requires more time than \MS, the difference remains within acceptable bounds, especially given the significant improvements in cartographic accuracy.

All experiments were executed using a Java implementation on a machine with an Intel\textsuperscript{\textregistered} Core\textsuperscript{TM} i9-13900H CPU @ 2.60 GHz and 16\,GB of RAM, running Windows 11 Home. Each algorithm was run in a single-threaded configuration to ensure consistent and fair timing measurements.

\end{appendix}

\end{document}